\documentclass[11pt,letter]{article}
\usepackage[section]{placeins}
\usepackage[centertags]{amsmath}
\usepackage{amssymb}
\usepackage{amsthm}
\usepackage{graphicx}
\usepackage{natbib}
\usepackage[nolists]{endfloat}
\usepackage{wasysym}
\usepackage{color}
\newcommand{\nothing}[1]{}

\newcommand{\beq}{\begin{equation}}
\newcommand{\eeq}{\end{equation}}
\newcommand{\bed}{\begin{displaymath}}
\newcommand{\eed}{\end{displaymath}}

\newlength{\defbaselineskip}
\newcommand{\setlinespacing}[1]%
           {\setlength{\baselineskip}{#1 \defbaselineskip}}
\newcommand{\doublespacing}{\setlength{\baselineskip}%
                           {1.8 \defbaselineskip}}

\def\b#1{\mbox{\boldmath $#1$}}    
\def\m#1{\mbox{#1}}                
\def\ml#1{\mbox{\scriptsize #1}} 
\def\ha#1{\mbox{$\hat{\b #1}$}}

\newcommand{\U}{\mathcal{U}}
\newcommand{\F}{\mathcal{F}}

\newtheorem{theorem}{Theorem}
\newtheorem{corollary}{Corollary}

\newtheorem{assume}{\rlap{A}\hskip
  .12in }

\begin{document}

\title{Calibration estimation in dual frame surveys}

\author{M. Giovanna Ranalli\thanks{Department of Political Sciences, Universit\`a degli Studi di Perugia, Italy, \texttt{giovanna.ranalli@stat.unipg.it}} \and Antonio Arcos\thanks{Department of Statistics and Operational Research, Universidad de Granada, Spain} \and Mar\'ia del Mar Rueda\thanks{Department of Statistics and Operational Research, Universidad de Granada, Spain} \and  Annalisa Teodoro\thanks{Department of Economics, Finance and Statistics, Universit\`a degli Studi di Perugia, Italy}}

\maketitle
\begin{abstract}
\noindent Survey statisticians make use of the available auxiliary information to improve estimates. One important example is given by calibration estimation, that seeks for new weights that are  close (in some sense) to the basic design weights and that, at the same time, match benchmark constraints on available auxiliary information. Recently, multiple frame surveys have gained much attention and became largely used by statistical agencies and private organizations to decrease sampling costs or to reduce frame undercoverage errors that could occur with the use of only a single sampling frame. Much attention has been devoted to the introduction of different ways of combining estimates coming from the different frames. We will extend the calibration paradigm, developed so far for one frame surveys, to the estimation of the total of a variable of interest in dual frame surveys as a general tool to include auxiliary information, also available at different levels. In fact, calibration allows us to handle different types of auxiliary information and can be shown to encompass as a special cases some of the methods already proposed in the literature. The theoretical properties of the proposed class of estimators are derived and discussed, a set of simulation studies is conducted to compare the efficiency of the procedure in presence of different sets of auxiliary variables. Finally, the proposed methodology is applied to data from the Barometer of Culture of Andalusia survey.
\end{abstract}

\noindent {\bf Keywords}: Auxiliary information, Kullback-Leibler distance, Raking ratio, Regression estimation, Survey Methodology.

\bibliographystyle{apalike}

\doublespacing

\section{Introduction}\label{sec:intro}

A main aim of survey statisticians is to obtain more accurate estimates, without increasing survey costs. Two popular tools to achieve this goal are $(i)$ the use of more than one population frame to select independent samples and $(ii)$ the use of auxiliary information either at the design or at the estimation stage. The use of more than one list of population units is important because a common practical problem in conducting sample surveys is that frames may be incomplete or out of date, so that resulting estimates may be seriously biased. Multiple frame surveys  are  useful when no single frame covers the whole
target population but the union of several available frames does, or when information
about a subgroup of particular interest comes only from an incomplete frame. They also have other advantages.
In fact, \citet{hartley1962multiple} introduces  dual frame surveys as a cost-saving device, showing that they
can often achieve the same precision as a single-frame survey at a much reduced cost. \citet{kalton1986sampling} suggest using two frames for sampling rare populations where even greater efficiencies can be obtained. Several estimators of the population total and mean have been proposed in the literature in dual frame surveys, usually classified, according to the level of frame information needed, as \emph{dual-frame} and \emph{single-frame} estimators.

On the other hand, the growing availability of information coming from census data, administrative registers and previous surveys provide a wide range of variables, concerning the population of interest, that are eligible to be employed as auxiliary information to increase efficiency in the estimation procedure. In this scenario, a very relevant example  is given by \textit{calibration estimation} that  adjusts basic design weights to account for auxiliary information and meet benchmark constraints on auxiliary variables population statistics \citep{Deville1992calibration}. \citet{sarndal2007calibration} provides an overview on developments in calibration estimation. In this paper, we will show how to extend calibration estimation to handle estimation from two frame surveys and how different types of auxiliary information can be easily integrated in the calibration process as benchmark constraints. Moreover, depending on the information available at the design stage, we show how to build calibration estimators under both the dual and the single frame approach. We will show that the proposed class of calibration estimators encompasses as particular cases some of the estimators already proposed in the literature. To show evidence of such connections, we will follow the minimum distance approach for calibration estimation, although using the instrumental variable approach is of course possible.

The paper is organized as follows. In Section \ref{sec:revdf} notation is introduced and those methods proposed in the literature to handle dual frame estimation are briefly reviewed. Then Section \ref{sec:caldualframe} illustrates the proposed class of calibration estimators by first dealing with the dual-frame approach and then moving, in Section \ref{sec:calsingleframe}, to the single-frame approach. The general form is provided and particular cases are derived according to relevant examples of auxiliary information. The theoretical  properties of the proposed estimators are investigated in an asymptotic framework adapted from that of \citet{isaki1982survey}. In addition, analytic and Jackknife variance estimators are proposed. Then, Section \ref{sec:sim} reports the results of an extensive simulation study run on a set of synthetic finite populations in which the performance of the proposed class of estimators is investigated for finite size samples. Section \ref{sec:appl} shows the application of the proposed estimation technique to data from the Barometer of Culture of Andalusia survey. Section \ref{sec:concl} provides some conclusions and directions for future research.

\section{Estimation in dual frame surveys}
\label{sec:revdf}

Consider a finite set of $N$ population units identified by the integers, $\mathcal{U}=\{1,\dots,k,\dots,N\}$, and let $A$ and $B$ be two sampling-frames, both can be incomplete, but it is assumed that together they cover the entire finite population. Let $\mathcal{A}$ be the set of population units in frame $A$ and $\mathcal{B}$ the set of population units in frame $B$. The population of interest, $\mathcal{U}$, may be divided into three mutually exclusive domains, $a=\mathcal{A}\cap\mathcal{B}^c, b=\mathcal{A}^c\cap\mathcal{B}$ and $ab=\mathcal{A}\cap\mathcal{B}$. Because the population units in the overlap domain $ab$ can be sampled in either survey or both surveys, it is convenient to create a duplicate domain $ba=\mathcal B\cap\mathcal A$, which is identical to $ab=\mathcal A\cap\mathcal B$, to denote the domain in the overlapping area coming from frame $B$.
Let $N$, $N_A$, $N_B$, $N_a$, $N_b$, $N_{ab}$, $N_{ba}$ be the number of population units in $\mathcal{U}$, $ \mathcal{A} $, $ \mathcal{B} $, $a$, $b$, $ab$, $ba$, respectively. It follows that $N_A=N_a+N_{ab}$, $ N_B=N_b+N_{ba}$ and $ N = N_a+N_b+N_{ab} = N_a+N_b+N_{ba}$.

Let $y$ be a variable of interest in the population and $y_k$ its value on  unit $k$, for $k=1, \ldots,N$. The entire set of population $y$ values is our finite population $\F$. The objective is to estimate the finite population total $Y=\sum_{k=1}^N y_k$ of $y$, that can be written as
\begin{equation}\label{eq:poptot}
Y=Y_a+\eta Y_{ab}+(1-\eta) Y_{ba} + Y_b,
\end{equation}
where $0\leq\eta\leq1$, and $Y_a=\sum_{k \in a}y_k$, $Y_{ab}=\sum_{k \in ab}y_k$, $Y_{ba}=\sum_{k \in ba}y_k$ and $Y_{b}=\sum_{k \in b}y_k$. Two probability samples $s_A$ and $s_B$ are drawn independently from frame $A$ and frame $B$ of sizes $n_A$ and $n_B$, respectively. Each design induces first-order inclusion probabilities $\pi_{Ak}$ and $\pi_{Bk}$, respectively, and sampling weights $d_{Ak}=1/\pi_{Ak}$ and $d_{Bk}=1/\pi_{Bk}$.
Units in $s_A$ can be divided as $s_A=s_a\cup s_{ab}$, where $s_a=s_A \cap a$ and $s_{ab}=s_A \cap (ab)$.  Similarly, $s_B=s_b\cup s_{ba}$,  where $s_b=s_B \cap b$ and $s_{ba}=s_B \cap (ba)$. Note that $s_{ab}$ and $s_{ba}$ are both from the same domain $ab$, but $s_{ab}$ is part of the frame $A$ sample and $s_{ba}$ is part of the frame $B$ sample. In this way, we have a sort of ``poststratified'' sample $s=s_a \cup s_{ab} \cup s_{ba} \cup s_b$ with ``poststratum" sample sizes $n_a$, $n_{ab}$, $n_{ba}$ and $n_b$. Note that $n_A=n_a+n_{ab}$ and $n_B=n_b+n_{ba}$ \citep[see][]{rao2010pseudo}.

The \citet{hartley1962multiple} estimator of $Y$ is given by
\begin{equation}
\label{eq:Hartleyest}
\hat{Y}_H(\eta)=\hat{Y}_a+\eta\hat{Y}_{ab}+(1-\eta)\hat{Y}_{ba}+ \hat{Y}_b,
\end{equation}
where  $\hat{Y}_a=\sum_{k\in s_a}d_{Ak}y_{k}$ is the Horvitz-Thompson estimator for the total of domain $a$ and similarly for the other domains.  If we let
\begin{equation*}
d_k^{\circ} = \left\{
\begin{array}{ll}
d_{Ak}                    & \text{if } k \in s_a\\
\eta d_{Ak}               & \text{if } k \in s_{ab}\\
(1-\eta) d_{Bk}           & \text{if } k \in s_{ba}\\
d_{Bk}                    & \text{if } k \in s_b\\
\end{array} \right.
\end{equation*}
then $\hat{Y}_H(\eta)=\sum_{k\in s}d_k^{\circ}y_k$. In the following, we will drop $\eta$ for ease of notation. Since each domain is estimated by its Horvitz-Thompson estimator, $\hat{Y}_H$ is an unbiased estimator of $Y$ for a given $\eta$. Since frames $A$ and $B$ are sampled independently, the variance of $\hat{Y}_H$  is given by
\begin{equation}
\label{eq:varHartleyest}
V(\hat{Y}_H)=V(\hat{Y}_a+\eta\hat{Y}_{ab})+V((1-\eta)\hat{Y}_{ba}+ \hat{Y}_b),
\end{equation}
where the first component of the right hand side is computed under $p_A(\cdot)$ (the sampling design in frame $A$) and the second one under $p_B(\cdot)$, and both are always understood conditional on the finite population $\F$. 

Choice of a value for $\eta$ has attracted much attention in literature; the value of $\eta$ that minimizes the variance in (\ref{eq:varHartleyest}) depends on unknown population variances and covariances and, when estimated from the data, it depends on the values of the variable of interest. This implies a need to recompute weights for every variable of interest $y$, which will  be inconvenient in practice for statistical agencies conducting surveys with numerous variables and lead to inconsistencies in  the estimates \citep[see][for a review]{lohr2009multiple}.

The estimator developed by \citet[FB]{fuller1972estimators} incorporates information regarding the estimation of $N_{ab}$ to improve over $\hat{Y}_H$, but has the drawback of not being a linear combination of $y$ values, unless using simple random sampling. \citet{skinner1996estimation} propose a modification of the estimator proposed by \citet{fuller1972estimators} for simple random sampling to handle complex designs. They introduce a pseudo maximum likelihood (PML) estimator that does not achieve optimality like the FB estimator, but it can be written as a linear combination of the observations and the same set of weights can be used for all variables of interest.

Recently, \citet{rao2010pseudo} extend the Pseudo-Empirical-Likelihood approach (PEL) proposed by \citet{wu2006pseudo} from one-frame surveys to dual-frame surveys following a stratification approach. They consider estimation of the population mean of $y$,
\begin{equation*}
\bar{Y}=W_a\bar{Y}_a+W_{ab}(\eta)\bar{Y}_{ab}+W_{ba}(\eta)\bar{Y}_{ba}+W_b\bar{Y}_b,
\end{equation*}
where $W_a=N_a/N$, $W_{ab}(\eta)=\eta N_{ab}/N$, $W_{ba}(\eta)=(1-\eta) N_{ab}/N$ and $W_b=N_b/N$, $\bar{Y}_{ab}=\bar{Y}_{ba}$, and again $\eta \in (0,1)$ is a fixed constant to be specified.  The PEL function takes the following expression:
\begin{eqnarray}\label{PELfunction}
l_D(p_{ak}, p_{abk}, p_{bak},p_{bk}) & = &
   n\Big[W_a\sum_{k\in s_a}\tilde{d}_{ak}\log(p_{ak})+W_{ab}(\eta)\sum_{k\in s_{ab}}\tilde{d}_{abk}\log(p_{abk})+\nonumber\\
   & &+W_{ba}(\eta)\sum_{k\in s_{ba}}\tilde{d}_{bak}\log(p_{bak})+W_{b}\sum_{k\in s_b}\tilde{d}_{bk}\log(p_{bk})\Big],
\end{eqnarray}
for all $k\in s$, where $n=n_A+n_B$, $\tilde{d}_{a_k}=d_{Ak}/\sum_{k\in s_a}d_{Ak}$, $\tilde{d}_{abk}=d_{Ak}/\sum_{k\in s_{ab}}d_{Ak}$, $\tilde{d}_{bk}=d_{Bk}/\sum_{k\in s_b}d_{Bk}$ and $\tilde{d}_{bak}=d_{Bk}/\sum_{k\in s_{ba}}d_{Bk}$. The four sets of probability measures in (\ref{PELfunction}) are found by maximizing the PEL function under the following normalizing constraints
$$
\sum_{k \in s_a}p_{ak}=1, \qquad \sum_{k \in s_{ab}}p_{abk}=1, \quad \sum_{k \in s_{ba}}p_{bak}=1, \qquad \sum_{k \in s_b}p_{bk}=1,
$$
and the constraint induced by the common domain mean $\bar{Y}_{ab}=\bar{Y}_{ba}$
\begin{equation}\label{PEL-constr-dualframe2}
\sum_{k \in s_{ab}}p_{abk}y_k = \sum_{k \in s_{ba}}p_{bak}y_k.
\end{equation}
The maximum PEL estimator of $\bar{Y}$ is then computed as \begin{equation}\label{PEL-est-dualframe}
\hat{\bar{Y}}_P=W_a\hat{\bar{Y}}_a+W_{ab}(\eta)\hat{\bar{Y}}_{ab}+
W_{ba}(\eta)\hat{\bar{Y}}_{ba}+W_b\hat{\bar{Y}}_b,
\end{equation}
where $\hat{\bar{Y}}_a=\sum_{k \in s_a}\hat{p}_{ak}y_k$, $\hat{\bar{Y}}_b=\sum_{k \in s_b}\hat{p}_{bk}y_k$ and $\hat{\bar{Y}}_{ab}=\sum_{k \in s_{ab}}\hat{p}_{abk}y_k=\hat{\bar{Y}}_{ba}$ because of constraint (\ref{PEL-constr-dualframe2}). Situations in which population domain sizes are not known are  sketched and the choice of $\eta$ is also discussed.

When inclusion probabilities in domain $ab$ are known for both frames, and not just for the frame from which the unit was selected,  \emph{single-frame} methods can be used that combine the observations into a single dataset and  adjust the weights in the intersection domain for multiplicity. In particular, observations from frame $A$ and frame $B$ are combined and the two samples drawn independently from $A$ and $B$ are considered as a single stratified sample over the three domains $a$, $b$ and $ab$. To adjust for multiplicity, the weights are defined as follows  for all units in frame $A$ and in frame $B$,
\begin{equation*}
d^\star_{k} = \left\{
\begin{array}{ll}
d_{Ak}                    & \text{if } k \in s_a\\
(1/d_{Ak}+1/d_{Bk})^{-1} & \text{if } k \in s_{ab}\cup s_{ba}\\
d_{Bk}                    & \text{if } k \in s_b\\
\end{array} \right..
\end{equation*}
{\color{black}Note that units in the overlap domain, which are expected to be selected a number of times given by $1/d_{Ak}+1/d_{Bk}$ have equal weights in frame $A$ and in frame $B$. } The estimator proposed by \citet{kalton1986sampling} is essentially an Horvitz-Thompson estimator for which
\begin{equation}
\label{KA-est}
\hat{Y}_S=\sum_{k \in s}d^\star_{k}y_k.
\end{equation}
Its variance is given by $ V(\hat{Y}_S) = V(\sum_{k\in s_A} d^{\star}_ky_k)+V(\sum_{k\in s_B} d^{\star}_ky_k)$, where the first component of the right hand side is computed under $p_A(\cdot)$ and the second one under $p_B(\cdot)$.   If $N_A$ and $N_B$ were known, the single-frame estimator $\hat{Y}_S$ could be adjusted using  raking ratio estimation \citep{bankier1986estimators,skinner1991efficiency}.

In the following section calibration estimation for dual frame surveys is introduced. We will first consider dual-frame methods and, then, to encompass situations in which auxiliary information is also in the form of inclusion probabilities for all units in both frames from both sampling design, single-frame methods will be considered as well (Section \ref{sec:calsingleframe}).

\section{Calibration estimation: dual-frame methods}
\label{sec:caldualframe}

In this section, we will show how to extend calibration estimation, as discussed in one frame surveys by \citet{Deville1992calibration}, to handle estimation from two frame surveys and how different types of auxiliary information can be easily integrated in the calibration process as benchmark constraints. Now, let $\b{x}_k=(x_{1k},\ldots,x_{pk})$  be the value taken on unit $k$ by a vector of auxiliary variables $\b x$ of which we assume to know the population total $\b{t}_x=\sum_{k=1}^N\b{x}_{k}$. This vector of totals may pertain  only $\mathcal{A}$, only $\mathcal{B}$, the entire population $\mathcal{U}$, or a combination of the three. We will first look at a general formulation of the problem, and then provide (relevant) examples of auxiliary vectors $\b x$.
Using the calibration paradigm, we wish to modify, as little as possible, basic Hartley weights $d_k^{\circ}$ to obtain new weights  $w_k^{\circ}$, for $k\in s$ to account for auxiliary information and derive a more accurate estimation of the total $Y$. A general dual-frame calibration estimator can be defined as
\begin{equation}\label{eq:calf}
\hat{Y}_{\ml{CAL}}=\sum_{k\in s} w_k^{\circ} y_k
\end{equation}
where $w_k^{\circ}$ is such that
\begin{equation}
\label{eq:calmindf}
\min \sum_{k \in s} G(w_k^{\circ},d_k^{\circ}) \qquad \text{s.t.} \qquad \sum_{k \in s} w_k^{\circ} \b{x}_k= \b{t}_x,
\end{equation}
where $G(w,d)$ is a distance measure satisfying the usual conditions required in the calibration paradigm \citep[see e.g.][Section 2]{Deville1992calibration}. Note that $\ha t_{xH}=\sum_{k \in s}d_k^{\circ} \b{x}_k$ is the Hartley estimator of $\b{t}_x$ for a given $\eta$. Then $w_k^{\circ}=d_k^{\circ}F(\b{x_k \b{\lambda}})$, where $F(u)=g^{-1}(u)$ and $g^{-1}(\cdot)$ denotes the inverse function of $g(w,d)=\partial G(w,d)/\partial w$. The vector $\b{\lambda}$ is determined using
$$
\phi_s(\b{\lambda})=\sum_{k \in s}d_k^{\circ}[F(\b{x}_k\b{\lambda})-1]\b{x}_k^T,
$$
so that $\phi_s(\b{\lambda})=\b{t}_x-\ha{t}_{xH}$.

Given the set of constraints, different calibration estimators are obtained by using different distance measures. In many instances, numerical methods are required to solve the the minimization problem in (\ref{eq:calmindf}). However, it is well known that, if we take the Euclidean (or $\chi^2$-statistic) type of distance function $G(w_k^{\circ},d_k^{\circ})=(w_k^{\circ}-d_k^{\circ})^2/2d_k^{\circ}$, equivalent to the \emph{linear} method in \citet{deville1993generalized}, we can obtain an analytic solution. In particular,
\begin{equation}
\label{weights}
w_k^{\circ}=d_k^{\circ}(1+\b{x}_k\b{\lambda})
\end{equation}
and, substituting this value in the calibration constraint in (\ref{eq:calmindf}), we obtain $\b{\lambda}=[\sum_{k \in s}d_k^{\circ}\b{x}_k^{T}\b{x}_k]^{-1}(\b{t}_x-\ha{t}_{xH})^{T}$. Substituting this value back into equation (\ref{weights}), the weights take the following form
\begin{equation}\label{eq:linweightsdf}
w_k^{\circ}=d_k^{\circ}\Big[1+(\b{t}_x-\ha{t}_{xH})\Big(\sum_{k \in s}d_k^{\circ}\b{x}_k^{T}\b{x}_k\Big)^{-1}\b{x}_k^{T}\Big].
\end{equation}
In this case, estimator $\hat{Y}_{\ml{CAL}}$ can be written as:
\begin{eqnarray*}
\hat{Y}_{\ml{CAL}} & = & \sum_{k\in s}w_k^{\circ}y_k=\hat{Y}_H+(\b{t}_x-\ha{t}_{xH})\ha{\beta}^{\circ} \nonumber \\
&=&\sum_{k=1}^N\b{x}_k\ha{\beta}^{\circ}+\sum_{k \in s}d_k^{\circ}(y_k-\b{x}_k\ha{\beta}^{\circ}),
\end{eqnarray*}
where $\ha{\beta}^{\circ}=(\sum_{k \in s}d_k^{\circ}\b{x}_k^{T}\b{x}_k)^{-1}(\sum_{k \in s} d_k^{\circ} \b{x}_k^Ty_k).$
This estimator takes the form of a \emph{generalized regression} type estimator for dual frame surveys and will be denoted by $\hat{Y}_{\ml{GREG}}$. These results are in line with those of calibration estimator in one frame surveys: Horvitz-Thompson estimators of $Y$ and $\b t_x$, and in the regression coefficient are here replaced by Hartley estimators.

Now, if we take as distance function $G(\cdot)$ the Kullback-Leibler divergence defined as
\begin{equation}
\label{cal-est-kl}
G(w_k^{\circ},d_k^{\circ})=-d_k^{\circ}\log(w_k^{\circ}/d_k^{\circ})+w_k^{\circ}-d_k^{\circ},
\end{equation}
that is  Case 4 distance examined in \citet{Deville1992calibration},  then $F(u)= 1/(1-u)$ and numerical methods are required. It can be noted that maximizing the PEL function in (\ref{PELfunction}) is equivalent to minimizing (\ref{cal-est-kl}) given the same set of starting weights and set of constraints. This equivalence was already noted in one frame surveys by \citet{deville2005talk}.

The calibration process induces a different final value for the weights which depends on both the  distance measure $G(\cdot,\cdot)$ used and the benchmark constraints applied. {\color{black}On the other hand, given a value for $\eta$, the final set of weights does not depend on the values of the variables of interest and can be, therefore, used for all variables of interest. When a value for $\eta$ is to be computed from the sample data, then it is essential to consider proposals based on estimators of $N_a$, $N_b$ and $N_{ab}$  as the one in, e.g.,  \citet{skinner1996estimation} so that it is the same for all variables of interest. } In the following, we consider some relevant examples of the form taken by the calibration estimator according to the auxiliary information available. Then, the theoretical properties are proven in Section \ref{sec:aspropdf}.

\subsection{$N_A$, $N_B$ and $N_{ab}$ all known}\label{sec:dfallknown}
Suppose that the dimension of the three sets $N_A$, $N_B$ and $N_{ab}$ is known. Then, we can build the auxiliary vector using domain membership indicator variables, i.e.
\begin{equation}
\label{aux-allknown}
\b{x}_k=\big(\delta_k(a),\delta_k(ab),\delta_k(ba),\delta_k(b)\big), \quad \text{for} \quad k=1,\ldots,N,
\end{equation}
where $\delta_k(a)=1$ if $k\in a$ and $0$ otherwise, $\delta_k(ab)=1$ if $k\in ab$ and $0$ otherwise, $\delta_k(ba)=1$ if $k\in ba$ and $0$ otherwise and $\delta_k(b)=1$ if $k\in b$ and $0$ otherwise. In order to have final weights that can be used directly to estimate population totals as in equation (\ref{eq:calf}), we will let  $\b{t}_x=(N_a,\eta N_{ab},(1-\eta)N_{ba},N_b)$ be the vector of known totals. In this case the calibration constraints are given by
\begin{equation}
\label{eq:constr4}
\sum_{k\in s_a}w_k^{\circ}=N_a, \quad \sum_{k\in s_{ab}}w_k^{\circ}=\eta N_{ab}, \quad \sum_{k\in s_{ba}}w_k^{\circ}=(1- \eta) N_{ba}, \quad \sum_{k\in s_{b}}w_k^{\circ}=N_{b},
\end{equation}
and the minimization problem has an analytic solution irrespective of the distance function employed. Such solution is given by
\begin{equation}\label{eq:weight-all-known}
w_k^{\circ} = \left\{
\begin{array}{ll}
d_{Ak}{N_a}/{\hat{N}_a} & \text{if } k \in s_a\\
\eta\, d_{Ak}   {N_{ab}}/{\hat{N}_{ab}}            & \text{if } k \in s_{ab}\\
(1-\eta)\, d_{Bk}  {N_{ba}}/{\hat{N}_{ba}}         & \text{if } k \in s_{ba}\\
d_{Bk}{N_{b}}/{\hat{N}_{b} }                   & \text{if } k \in s_b\\
\end{array} \right.,
\end{equation}
{\color{black}where $\hat{N}_a=\sum_{k\in s_a}d_{Ak}$, $\hat{N}_{ab}=\sum_{k\in s_{ab}}d_{Ak}$, $\hat{N}_{ba}=\sum_{k\in s_{ba}}d_{Bk}$ and $\hat{N}_b=\sum_{k\in s_b}d_{Bk}$. } Note that these weights provide H\'ajek type estimators for each domain and mirror the result provided in \cite{deville1993generalized} when dealing with  the calibration estimator in case auxiliary information consists of known cell counts in a frequency table. \cite{deville1993generalized} denote this case as \emph{complete post-stratification}, that is when all the domain sizes are known and used for calibration. {\color{black}{Note that, given that we are estimating totals in domains using ratio type estimators, the sample size of the domains is important to avoid the introduction of possible bias in the final estimates.}}

\subsection{$N_A$, $N_B$ known and $N_{ab}$ unknown}
\label{sec:NabunknownDF}
Following the terminology of \cite{deville1993generalized}, we call the  case treated in this section as \emph{incomplete post-stratification} and we mean that not all the domain sizes are known, in particular we know only the size of frame $A$ and of frame $B$, but we don't known the size of the overlap domain $ab$. In this case, for $k=1,\ldots,N$, we can write the vector $\b{x}$ of auxiliary information as:
\begin{equation}
\label{aux-vect-Nabunknown}
\b{x}_k=(\delta_k(a)+\delta_k(ab)+\delta_k(ba),\delta_k(b)+\delta_k(ab)+\delta_k(ba)).
\end{equation}
The vector of known totals in this case is $\b{t}_x=(N_A,N_B)$ and we have the following calibration constraints
\begin{align}
&\sum_{k\in s_a}w_k^{\circ}+\sum_{k\in s_{ab}}w_k^{\circ}+\sum_{k\in s_{ba}}w_k^{\circ}=N_A \nonumber\\
&\sum_{k\in s_b}w_k^{\circ}+\sum_{k\in s_{ab}}w_k^{\circ}+\sum_{k\in s_{ba}}w_k^{\circ}=N_B,\nonumber
\end{align}
in which we, in some sense,  calibrate on the margins. Final calibration weights are no longer independent from the distance function used and it is not possible to obtain an analytical expression unless we use the Euclidean distance. In this latter case we obtain the following estimator of $N_{ab}$:
\begin{equation}\label{Nab-est}
\hat{N}_{ab}^w=\hat{N}_{ab,H}\frac{\hat{N}_a N_B+\hat{N}_b N_A - \hat{N}_a\hat{N}_b}
{\hat{N}_a \hat{N}_B+\hat{N}_b \hat{N}_A - \hat{N}_a\hat{N}_b},
\end{equation}
where $\hat{N}_{ab,H}=\eta \hat{N}_{ab} + (1-\eta) \hat{N}_{ba}$. We can note how the calibration procedure adjusts the Hartley estimator of $N_{ab}$ accounting for auxiliary information.

\cite{rao2010pseudo} also consider the case in which $N_{ab}$ is unknown. However, they do not estimate it from within the maximum PEL procedure, but they first estimate it by $\hat{N}_{ab,P}=\hat{\theta}\hat{N}_{ab}+(1-\hat{\theta})\hat{N}_{ba}$, where $\hat{\theta}=v(\hat{N}_{ba})/\{v(\hat{N}_{ab})+v(\hat{N}_{ba})\}$ and $v$ denotes variance estimates. Then, they  take a \emph{pseudo-complete post stratification} approach by suitably modifying the likelihood function.

\subsection{{\color{black}Population totals for group membership indicators are known}}
\label{sec:groups}
{\color{black}
Let the population $\mathcal{U}$ be divided into $H$ mutually exclusive groups $\U_h$, for $h=1,\ldots,H$ such that $\bigcup_{h=1}^H \U_h=\U$ and let $\delta_k(h)$ be the indicator variable that takes value 1 if unit $k\in \U_h$ and 0 otherwise, for $k=1,\ldots,N$ and $h=1,\ldots,H$. Then, $\sum_{k=1}^N\delta_k(h)=N_h$ and $\sum_{h=1}^{H}N_h=N$. Now, consider the situation in which we know the population total of such indicator variables for each of the four domains, i.e. $N_{a,h}=\sum_{k\in a}\delta_k(h)$, $N_{ab,h}=\sum_{k\in ab}\delta_k(h)$, $N_{ba,h}=\sum_{k\in ba}\delta_k(h)=N_{ab,h}$, $N_{b,h}=\sum_{k\in b}\delta_k(h)$, for $h=1,\ldots,H$. Note that $N_{a,h}=\sum_{k\in a}\delta_k(h)=\sum_{k=1}^N\delta_k(a)\delta_k(h)$ and similarly for the other cases.  In practice, this would mean that we know, say the number of  units for each of $H$ age-sex groups in the population for each of the four domains. This amount of auxiliary information of course implies that we also know the dimension of the three sets $N_A$, $N_B$ and $N_{ab}$ considered in the Section \ref{sec:dfallknown}. Indeed, that is a special case of the present one. 

In this case the vector of  auxiliary variables is defined for $k=1,\ldots,N$ by
$$\b{x}_k=\{(\delta_k(a)\delta_k(h),\delta_k(ab)\delta_k(h),\delta_k(ba)\delta_k(h),\delta_k(b)\delta_k(h)\}_{h=1,\ldots,H} $$
and the vector of known totals is set to be $\b{t}_x=\{(N_{a,h},\eta N_{ab,h},(1-\eta)N_{ba,h},N_{b,h})\}_{h=1,\ldots,H}$. As in Section \ref{sec:dfallknown} the minimization problem has an analytic solution irrespective of the distance function employed. Such solution is given by 
\begin{equation}\label{eq:groupsh}
w_k^{\circ} = \left\{
\begin{array}{ll}
d_{Ak}{N_{a,h}}/{\hat{N}_{a,h}} & \text{if } k \in \{s_a \cap \U_h\}\\
\eta\, d_{Ak}   {N_{ab,h}}/{\hat{N}_{ab,h}}            & \text{if } k \in \{s_{ab} \cap \U_h\}\\
(1-\eta)\, d_{Bk}  {N_{ba,h}}/{\hat{N}_{ba,h}}         & \text{if } k \in \{s_{ba} \cap \U_h\}\\
d_{Bk}{N_{b,h}}/{\hat{N}_{b,h} }                   & \text{if } k \in \{s_b \cap \U_h\}\\
\end{array} \right. \m{for $h=1,\ldots,H$,}
\end{equation}
where $\hat{N}_{a,h}=\sum_{k\in s_a}d_{Ak}\delta_k(h)$ and similarly for the other size estimators. This is another case of complete post-stratification. The final estimator will be more efficient than the Hartley estimator as much as groups collect units with a similar value of the variable of interest. 

When, on the other side, we only know the population total in frame $A$ and in frame $B$, i.e. we do not know the distribution for the intersection domain $ab$, then we are again in a situation of incomplete post-stratification, like that of Section \ref{sec:NabunknownDF}. Here, $$\b{x}_k=\{[\delta_k(a)+\delta_k(ab)+\delta_k(ba)]\delta_k(h),[\delta_k(b)+\delta_k(ab)+\delta_k(ba)]\delta_k(h)\}_{h=1\ldots,H}$$ and $\b{t}_x=\{(N_{A,h},N_{B,h})\}_{h=1\ldots,H}$. We have an analytic solution for the form of the weights only for the Euclidean distance case, but it does not take a  simple tractable form as that considered in Section \ref{sec:NabunknownDF}. A similar situation arises also when, as in the case considered later in the application (Section \ref{sec:appl}), we do not know the distribution for, say, age-sex groups, but we know only the total for age and the total for sex, in each of the two frames $A$ and $B$. This is another example of incomplete post-stratification, that employs a form of raking (depending on the distance function employed) to obtain the final set of weights (see also examples in Section \ref{sec:calsingleframe}). }

\subsection{{\color{black}$N_A$, $N_B$, $N_{ab}$ known and $X_A$ known}}
\label{sec:xadf}

Suppose that we know not only the frame sizes $N_A$, $N_B$ and $N_{ab}$, but, also the population total of an auxiliary numerical variable $x_A$ correlated to the study variable $y$ and relative to frame $A$, whose total is $X_A=\sum_{k \in \mathcal{A}} x_{Ak}$. In this case the vector of  auxiliary variables is defined for $ k=1\,\ldots,N$ by
$$\b{x}_k=(\delta_k(a),\delta_k(ab),\delta_k(ba),\delta_k(b),[\delta_k(a)+\delta_k(ab)+\delta_k(ba)] x_{Ak}) $$
and the calibration constraints are those in (\ref{eq:constr4}) plus
\begin{equation}\label{eq:constr5}
\sum_{k\in s_a}w_k^{\circ}x_{Ak}+\sum_{k\in s_{ab}}w_k^{\circ}x_{Ak}+\sum_{k\in s_{ba}}w_k^{\circ}x_{Ak}=X_A.
\end{equation}
Again, it is not possible to obtain an analytic expression for the calibration weights unless we use the Euclidean distance for the Lagrange function. It can be shown that, in this case, the calibrated weights for  $k\in s_a$ are such that 
\begin{equation}\label{weights-a-with-XA}
w_k^{\circ}=d_{Ak}\left[\frac{N_a}{\hat{N}_a}+\lambda(\frac{\hat{X}_a}{\hat{N}_a}-x_{Ak})\right],
\end{equation}
where $\lambda$ is the Lagrange multiplier for the last constraint in (\ref{eq:constr5}) given by
$$\lambda=\frac{X_A-\hat{X}_{A,\ml{H\'aj}}}{\hat{S}_{a,x}^2+\eta\hat{S}_{ab,x}^2+(1-\eta)\hat{S}_{ba,x}^2}
$$
where $\hat{X}_{A,\ml{H\'aj}}$ is a Hartley type estimator in which each component is estimated using the H\'ajek estimator,   $\hat{S}_{a,x}^2=\sum_{k \in s_a}d_{ak}(x_{Ak}-\hat{X}_a/\hat{N}_a)^2$ and similarly for $\hat{S}_{ab,x}^2$ and $\hat{S}_{ba,x}^2$. Calibrated weights $w_k^{\circ}$ for $k\in s_{ab}$ and for $k\in s_{ba}$ are similar to those in (\ref{weights-a-with-XA}) but with quantities referred to the appropriate domain, while weights for $k\in s_b$ are  the same as in (\ref{eq:weight-all-known}). With such weights, the resulting calibration estimator resembles a combined regression estimator; in fact \begin{equation*}\label{est-XA}
\hat{Y}_{\ml{CAL}}=\hat{Y}_{\ml{H\'aj}}+(X_A-\hat{X}_{A,\ml{H\'aj}})\hat{\beta}_A
\end{equation*}
where $\hat{Y}_{\ml{H\'aj}}$ is the Hartley estimator of $Y$  in which each component is estimated by its H\'ajek estimator, while
\begin{equation*}
\hat{\beta}_A=\frac{\hat{S}_{a,xy}+\eta\hat{S}_{ab,xy}+(1-\eta)\hat{S}_{ba,xy}}
{\hat{S}_{a,x}^2+\eta \hat{S}_{ab,x}^2+(1-\eta) \hat{S}_{ba,x}^2},
\end{equation*}
with $\hat{S}_{a,xy}=\sum_{k \in s_a}d_{ak}(x_{Ak}-\hat{X}_a/\hat{N}_a)(y_{k}-\hat{Y}_a/\hat{N}_a)$ and similarly for $\hat{S}_{ab,xy}$ and $\hat{S}_{ba,xy}$.

\subsection{{\color{black}Other examples}}

{\color{black}The cases previously discussed  are only a few examples of the very many possible ones that can be treated with calibration. The calibration approach is very flexible and can also handle both indicator and numerical variables simultaneously. Next we provide some details on how to construct the auxiliary vector and the vector of control totals for other  interesting cases in practice; some of these cases  will be used in the simulation study and in the application.

\begin{description}
\item[$N_A$, $N_B$, $N_{ab}$ known and $X$ known.]
Suppose that we know the frame sizes $N_A$, $N_B$ and $N_{ab}$, and let the population total of an auxiliary numerical variable  be available for the whole population $X= \sum_{k=1}^N x_{k}$ and not only for frame $A$ as in the previous section. The auxiliary vector is thus $\b{x}_k=(\delta_k(a),\delta_k(ab),\delta_k(ba),\delta_k(b), x_{k}) $
and the calibration constraints are those in  (\ref{eq:constr4}) plus
$\sum_{k\in s}w_k^{\circ}x_{k}=X.$

\item[$N_A$, $N_B$, known and $X_A$ and $Z_B$ known.]
Suppose that we know the frame sizes $N_A$, $N_B$ and the population total of an auxiliary numerical variable  $x_A$  relative to frame $A$, whose total is $X_A=\sum_{k \in \mathcal{A}} x_{Ak}$ and  the population total of another auxiliary numerical variable  $z_B$ relative to frame $B$, whose total is $Z_B=\sum_{k \in \mathcal{B}} z_{B}$.
The auxiliary vector is
$$\b{x}_k=(\delta_k(a)+\delta_k(ab)+\delta_k(ba),\delta_k(b)+\delta_k(ab)+\delta_k(ba), [\delta_k(a)+\delta_k(ab)+\delta_k(ba)] x_{Ak}, [\delta_k(b)+\delta_k(ab)+\delta_k(ba)] z_{Bk})$$ and the vector of known totals in this case is $\b{t}_x=(N_A,N_B,X_A,Z_B)$, which allows us to write the following calibration constraints
\begin{align}
&\sum_{k\in s_a}w_k^{\circ}+\sum_{k\in s_{ab}}w_k^{\circ}+\sum_{k\in s_{ba}}w_k^{\circ}=N_A \nonumber\\
&\sum_{k\in s_b}w_k^{\circ}+\sum_{k\in s_{ab}}w_k^{\circ}+\sum_{k\in s_{ba}}w_k^{\circ}=N_B,\nonumber\\
&\sum_{k\in s_a}w_k^{\circ}x_{Ak}+\sum_{k\in s_{ab}}w_k^{\circ}x_{Ak}+\sum_{k\in s_{ba}}w_k^{\circ}x_{Ak}=X_A \nonumber\\
&\sum_{k\in s_b}w_k^{\circ}z_{Bk}+\sum_{k\in s_{ab}}w_k^{\circ}z_{Bk}+\sum_{k\in s_{ba}}w_k^{\circ}z_{Bk}=Z_B.
\end{align}

\item[$N_A$, $N_B$, $N_{ab}$ known and $X_A$, $X_B$ known.]
When we know the frame sizes $N_A$, $N_B$ and $N_{ab}$ and the population totals of the same auxiliary variable $x$ in the two frames $X_A$ and $X_B$, the auxiliary vector is $$\b{x}_k=(\delta_k(a),\delta_k(ab),\delta_k(ba),\delta_k(b),
[\delta_k(a)+\delta_k(ab)+\delta_k(ba)] x_{k}, [\delta_k(b)+\delta_k(ab)+\delta_k(ba)] x_{k})$$ and the vector of known totals in this case is $\b{t}_x=(N_a,\eta N_{ab},(1-\eta)N_{ba},N_b,X_A,X_B)$.

\end{description}
}

\subsection{Asymptotic properties of $\hat{Y}_{\ml{CAL}}$}
\label{sec:aspropdf}
To show the asymptotic properties of the general calibration estimator we adapt and place ourselves in the asymptotic framework of \cite{isaki1982survey}, in which the dual-frame finite population $\mathcal{U}$ and the sampling designs $p_A(\cdot)$ and $p_B(\cdot)$ are embedded into a sequence of such populations and designs indexed by $N$, $\{\mathcal{U}_N,p_{A_N}(\cdot),p_{B_N}(\cdot)\}$, with $N\rightarrow\infty$. We will assume therefore, that $N_{A_N}$ and $N_{B_N}$ tend to infinity and that also $n_{A_N}$ and $n_{B_N}$ tend to infinity as $N\rightarrow\infty$. We will further assume that $N_a>0$ and $N_b>0$. In addition $n_{A_N}/n_N \rightarrow c_1 \in (0,1)$, where $n_N=n_{A_N}+n_{B_N}$, $N_a/N_A\rightarrow c_2 \in (0,1)$, $N_b/N_B\rightarrow c_3 \in (0,1)$ as $N\rightarrow\infty$. Subscript $N$ may be dropped for ease of notation, although all limiting processes are understood as $N\rightarrow\infty$. Stochastic orders $O_p(\cdot)$ and $o_p(\cdot)$ are with respect to the aforementioned sequences of designs. The constant $\eta \in (0,1)$ is kept fixed over repeated sampling. In order to prove our results, we make the following technical assumptions.
\begin{assume}\label{as:B}
Let $\b{B}_U=(\sum_{k=1}^N  \b{x}_k^{T}\b{x}_k)^{-1}\sum_{k=1}^N\b{x}_k ^Ty_k$. Assume that $\b{B}=\lim_{N\rightarrow\infty} \b{B}_U$ exists; the distribution of $\b{x}_k$ and of $y_k$, and the sampling designs are such that $\sum_{k=1}^N \b{x}_k^T \b{x}_k$ is consistently estimated by $\sum_{k \in s} d_k^{\circ}\b{x}_k^T \b{x}_k$ and $\sum_{k=1}^N \b{x}_k^T y_k$ is consistently estimated by $\sum_{k \in s}d_k^{\circ}\b{x}_k^T y_k$.
\end{assume}

\begin{assume}\label{as:poscov}
 The limiting design covariance matrix of the normalized Hartley estimators,
    \begin{equation*}
    \b{\Sigma}=
    \begin{bmatrix}
      \Sigma_{yy}            & \b{\Sigma}_{xy} \\
      \b{\Sigma}_{xy}^{T} & \b{\Sigma}_{xx}\\
    \end{bmatrix}
    =\lim_{N\rightarrow\infty} \frac{n_N}{N^2}
    \begin{bmatrix}
      V(\hat{Y}_H)            & \b{C}(\ha{t}_{xH},\hat{Y}_H) \\
      \b{C}(\ha{t}_{xH},\hat{Y}_H)^{T} & \b{V}(\ha{t}_{xH})\\
    \end{bmatrix}
    \end{equation*}
    is positive defined.
\end{assume}

\begin{assume}\label{as:clt}
The normalized Hartley estimators of $\b{t}_x$ and $Y$ are such that a central limit theorem holds:
    \begin{equation*}
    \frac{\sqrt{n_N}}{N}
    \begin{bmatrix}
      \sum_{k \in s} d_k^{\circ} y_k - Y \\
      \sum_{k \in s} d_k^{\circ} \b{x}_k^{T} - \b{t}_x^T\\
    \end{bmatrix}
    \to^{\hskip -0.125in \mathcal{L}} N(\b{0},\b{\Sigma}).
    \end{equation*}
\end{assume}

\begin{assume}\label{as:estvar}
The estimated covariance matrix for the Hartley estimator is design consistent in the sense that
    \begin{equation*}
    \frac{n_N}{N^2}
    \begin{bmatrix}
      v(\hat{Y}_H)            & \b{c}(\ha{t}_{xH},\hat{Y}_H) \\
      \b{c}(\ha{t}_{xH},\hat{Y}_H)^{T} & \b{v}(\ha{t}_{xH})\\
    \end{bmatrix}
    -\b{\Sigma}=o_p(1),
    \end{equation*}
    where $v(\hat{Y}_H)=v(\hat{Y}_a+\eta\hat{Y}_{ab})+v((1-\eta)\hat{Y}_{ba}+\hat{Y}_b)$ and similarly for the others.

\end{assume}

We will first state the properties of $\hat{Y}_{\ml{CAL}}$ for the Euclidean distance, i.e.  $\hat{Y}_{\ml{GREG}}$, and then show the convergence for a general distance function. The following theorem shows that $\hat{Y}_{\ml{GREG}}$ is design consistent, and provides its asymptotic distribution.

\begin{theorem} \label{the:cons}
Under assumptions {A\ref{as:B}--A\ref{as:clt}}, $\hat{Y}_{\ml{GREG}}$ is design $\sqrt{n_N}$-consistent for $Y$ in the sense that,
\begin{equation*}
\hat{Y}_{\ml{GREG}}-Y=O_p(Nn_N^{-1/2})
\end{equation*}
and has the following asymptotic distribution
$$
\frac{\hat{Y}_{\ml{GREG}}-Y}{\sqrt{V_{\infty}(\hat{Y}_{\ml{GREG}})}} \to^{\hskip -0.125in \mathcal{L}} N(0,1)
$$
where $V_{\infty}(\hat{Y}_{\ml{GREG}})=V(\hat{t}_{eH})$ and $\hat{t}_{eH}=\sum_{k\in s}d_k^{\circ}e_k$ is the Hartley estimator of the population total of the ``census''-level residuals $e_k=y_k-\b{x}_k \b{B}_U$.

\end{theorem}

\noindent \textbf{Proof.} See the Appendix.

\qed

A design unbiased variance estimator is available for the Horvitz-Thompson estimator for many designs, and therefore for the Hartley estimator for a given $\eta$. The following theorem shows that, in these cases, it is possible to construct  a design consistent estimator for the variance of the asymptotic distribution $V_{\infty}(\hat{Y}_{\ml{GREG}})$ obtained in Theorem \ref{the:cons}.

\begin{theorem} \label{the:estvar}
Let $\hat{e}_{k}=y_k-\b{x}_k \ha\beta^{\circ}$. Then, under assumptions {A\ref{as:B}, A\ref{as:poscov}}  and {A\ref{as:estvar}}
\begin{align*}
v(\hat{Y}_{\ml{GREG}})= v(\hat{t}_{\hat{e}H}) &=  v\Big(\sum_{k \in s_a}d_k\hat{e}_k+\eta\sum_{k \in s_{ab}}d_k\hat{e}_k\Big)+v\Big((1-\eta)\sum_{k \in s_{ba}}d_k\hat{e}_k+\sum_{k \in s_b}d_k\hat{e}_k\Big)=\nonumber\\
&=V(\hat{t}_{eH})+o_p(N^2n_N^{-1}).
\end{align*}
\end{theorem}

\noindent \textbf{Proof.} See the Appendix.

\qed

From Theorem \ref{the:estvar} we can derive an asymptotic distribution result using the  estimated variance as stated in the following corollary.

\begin{corollary} \label{cor:pivot}
Under assumptions {A\ref{as:B}--A\ref{as:estvar}}, $\hat{Y}_{\ml{GREG}}$ is such that
\begin{equation*}
\frac{\hat{Y}_{\ml{GREG}}-Y}{\sqrt{v(\hat{Y}_{\ml{GREG}})}}\to^{\hskip -0.125in \mathcal{L}}N(0,1).
\end{equation*}

\end{corollary}

Now we establish the asymptotic equivalence between $\hat{Y}_{\ml{CAL}}$ and $\hat{Y}_{\ml{GREG}}$. To this end, we further make the following assumptions \citep[see Section 2][]{Deville1992calibration}.
\begin{assume}\label{as:phi}
$\phi_s(\b{\lambda})$ is defined on $C=\bigcap_{k=1}^N\{\b{\lambda}:\b{x}_k\b{\lambda} \in \m{Im}_k(d_k^{\circ})\}$. $C$ is an open neighborhood of \b{0}.
\end{assume}
\begin{assume}\label{as:boundx}
As $N\rightarrow\infty$, $\max ||\b{x}_k||=M<\infty$, $k=1,\ldots,N$,  and $\max F_k^{\prime\prime}(0)=M^\prime<\infty$, where $F_k^{\prime\prime}(\cdot)$ is the second derivative of $F_k(\cdot)$.
\end{assume}

\begin{theorem} \label{the:cal}
Under assumptions {A\ref{as:B}--A\ref{as:clt}}  and {A\ref{as:phi}--A\ref{as:boundx}}
\begin{equation*}
\hat{Y}_{\ml{CAL}}-\hat{Y}_{\ml{GREG}}=O_p(Nn_N^{-1}).
\end{equation*}
\end{theorem}

\noindent \textbf{Proof.} See the Appendix.

\qed

\begin{corollary}\label{cor:cal}
Under {A\ref{as:B}--A\ref{as:boundx}} $\hat{Y}_{\ml{CAL}}$ is such that
\begin{equation*}
\frac{\hat{Y}_{\ml{CAL}}-Y}{\sqrt{v(\hat{Y}_{\ml{GREG}})}}\to^{\hskip -0.125in \mathcal{L}}N(0,1).
\end{equation*}
\end{corollary}
\section{Calibration estimation: single-frame methods}
\label{sec:calsingleframe}

In those situations in which we know the inclusion probability of the units in the sample under both sampling designs, then we can account for it in a calibration framework employing  \emph{single-frame} estimators \citep{kalton1986sampling,bankier1986estimators,skinner1991efficiency}. The calibration estimator in this single-frame approach is given by ${\hat{Y}}_{\ml{CAL}}^{\ml{S}}= \sum_{k\in s}{w}^{\star}_{k} y_k$ where weights  ${w}^{\star}_{k}$ are such that $$ \min \sum_{k\in s} G(w^{\star}_{k}, d_k^{\star}) \quad \m{s.t.} \quad \sum_{k\in s }w_k^{\star}\b{x}_k = \b{t}_x.$$
A general solution to the minimization problem is given by $w^{\star}_k=d^{\star}_kF(\b x_k\b\lambda)$. Note the only difference with equation (\ref{eq:calmindf}) is the starting basic design weight. {\color{black} Note that calibration can handle the case in which $(1/d_{Ak}+1/d_{Bk})\geq 1$ for some units $k$ and, therefore, the basic weights are smaller than 1. }

If we take the Euclidean distance function, the calibration weights obtained from the minimization procedure are given by
$w_k^{\star}=d_k^{\star}(1+\b{x}_k \b\lambda)$ with $\b \lambda = (\sum_{k\in s} d_k^{\star} \b{x}_k^T \b{x}_k)^{-1}( \b{t}_x -    \ha{t}_{xS} )^T$ and $  \ha{t}_{xS}=  \sum_{k\in s}d_k^{\star} \b{x}_k$, i.e. the single-frame estimator for the total $\b{t}_x$. As expected, the resulting calibration estimator takes a generalized regression estimator  form, given by
\begin{align*}
\hat{Y}_{\ml{GREG}}^{\ml{S}} & = \sum_{k\in s} d_k^{\star} y_k + (  \b{t}_x-   \ha{t}_{xS} )\Big(\sum_{k\in s} {d_k^{\star}}\b{x}_k^T \b{x}_k\Big)^{-1}  \sum_{k\in s}d_k^{\star} \b{x}_k^T  y_k \\
& = \hat{Y}_{S} +  (  \b{t}_x-   \ha{t}_{xS} ) \ha\beta^{\star},
\end{align*}
where  $ \ha\beta^{\star}=(\sum_{k\in s} d_k^{\star} \b{x}_k^T \b{x}_k)^{-1}  \sum_{k\in s}d_k^{\star}  \b{x}_k^T y_k$. Under assumptions in all similar to those of Section \ref{sec:aspropdf}, concerning consistency of the single-frame estimator instead of the Hartley estimator, $\hat{Y}_{\ml{CAL}}^{\ml{S}}$ can be proven to be a consistent estimator, to be asymptotically equivalent to $\hat{Y}_{\ml{GREG}}^{\ml{S}}$ and, therefore, to share its asymptotic distribution. In  particular, it can be easily shown that the variance of their asymptotic distribution is given by $V_{\infty}(\hat{Y}_{\ml{GREG}}^{\ml{S}})=V(\hat{t}_{eS})=V(\sum_{k\in s}d^{\star}_ke_k)$ and it can be consistently estimated using $v(\hat{t}_{\hat{e}S})=v(\sum_{k\in s}d^{\star}_k\hat{e}_k)$, where $\hat{e}_k=y_k-\b{x}_k\ha\beta^{\star}$. Of course here, as in the dual frame context previously explored,  variance estimators alternative to the one considered here based on the linearization technique and proposed in the literature to estimate the variance of the calibration estimator can be considered as well, once the set of basic design weights are properly adjusted for (e.g. those based on resampling methods or on empirical likelihood methods). We will consider Jackknife later in Section \ref{sec:jack}.

Given that in the single-frame approach each unit in the overlap domain has a weight that accounts for the expected number of times it can be selected in the sample, care should be placed in the definition of the auxiliary variable vector. In particular, in the case  $N_A$, $N_B$ and $N_{ab}$  are all known, $\b{x}_k= (\delta_k(a),\delta_k(ab)+\delta_k(ba),\delta_k(b)),$ and, therefore, $ \b{t}_x=(N_a,N_{ab},N_{b}$). As in Section \ref{sec:dfallknown}, the final solution does not depend on the choice of the distance function and calibrated weights take the H\'ajek form
\begin{equation*}
w_k^{\star} = \left\{
\begin{array}{ll}
d_{Ak}{N_a}/{\hat{N}_a} & \text{if } k \in s_a\\
(1/d_{Ak}+1/d_{Bk})^{-1}  {N_{ab}}/{\hat{N}_{abS}}            & \text{if } k \in s_{ab}\cup s_{ba}\\
d_{Bk}{N_{b}}/{\hat{N}_{b} }                   & \text{if } k \in s_b\\
\end{array} \right..
\end{equation*}
where $\hat{N}_{abS}=\sum_{k\in s_{ab}\cup s_{ba}}(1/d_{Ak}+1/d_{Bk})^{-1}$. Similarly, if $N_A$, $N_B$, $N_{ab}$, and $X_A$  are known, then $\b{x}_k= (\delta_k(a),\delta_k(ab)+\delta_k(ba),\delta_k(b),[\delta_k(a)+
 \delta_k(ab)+ \delta_k(ba)]x_{Ak} ),$ and  $\b{t}_x=(N_a,N_{ab},N_{b}, X_A$).

When, on the other hand, only  $N_A$, $N_B$ are known, an interesting equivalence arises. In this case the auxiliary vector is defined as in (\ref{aux-vect-Nabunknown}) and final weights depend on the distance function employed.  If we consider the Case 2 distance proposed  in    \citet{Deville1992calibration}, i.e.  $$G(w_k^{\star},d_k^{\star})=w_k^{\star}\log(w_k^{\star}/d_k^{\star})-w_k^{\star}
+d_k^{\star},$$ and the particular case of simple random sampling in both frames, we obtain that $\hat{N}_{ab}^w = \sum_{k\in s_{ab}\cup s_{ba}} w_k^\star= \hat{N}_{ab}^{RR}$, where $\hat{N}_{ab}^{RR}$ is the overlap dimension estimator obtained by \cite{skinner1991efficiency} using Raking Ratio as the smallest root of the quadratic equation
\begin{equation*}
\hat{N}_{abS}\,t^2-[\hat{N}_{abS}(N_A+N_B)+(\hat{N}_{a}\hat{N}_{ab})n_an_b]t+\hat{N}_{abS}N_AN_B=0.
\end{equation*}
In this case the single frame calibration estimator provides a simple tool to extend such  Raking Ratio estimator to general sampling designs by simply plugging in different basic design weights $d_k^\star$, and to more composite auxiliary information settings.

\section{Jackknife estimation of variance}
\label{sec:jack}

In this section we explore the possibility of using Jackknife to estimate the variance of the proposed calibration estimators \citep[see e.g.][for an introduction to Jackknife methods]{wolter:2007}. Dual-frame or single-frame calibration estimators will be denoted by $\hat{Y}_{c}$ for short in this section.

If we consider a non stratified design, the Jackknife estimator for the variance of $\hat{Y}_{c}$ may be given by
\begin{equation} \label{eq:jack}
 \displaystyle
v_J (\hat{Y}_{c}) =\frac{n_{A}-1}{n_{A}} \sum_{i\in s_A} (\hat{Y}_{c}^{A}(i) -\overline{Y}_{c}^{A})^2 + \frac{n_{B}-1}{n_{B}} \sum_{j\in s_B} (\hat{Y}_{c}^{B}(j) -\overline{Y}_{c}^{B})^2
\end{equation}
where $\hat{Y}_{c}^{A}(i)$ is the value taken by  estimator  $\hat{Y}_{c}$ after dropping unit $i$ from $s_A$ and $\overline{Y}_{c}^{A}$ is the average of $\hat{Y}_{c}^{A}(i)$ values; $\hat{Y}_{c}^{B}(j)$ and  $\overline{Y}_{c}^{B}$ are  defined similarly. This Jackknife estimator of the  variance is conservative (upward biased) in  finite populations when  sampling without replacement \citep[see][Section 4.3.4]{wolter:2007}. To overcome this issue, an approximate finite-population correction is employed. Then, the new Jackknife estimator of variance $v_J^* (\hat{Y}_{c})$ is obtained by replacing $\hat{Y}_{c}^{A}(i)$ in (\ref{eq:jack}) with  $\hat{Y}_{c}^{A *}(i)= \hat{Y}_{c}+\sqrt{1-\overline{\pi}_A} (\hat{Y}_{c}^{A}(i) -\hat{Y}_{c})$, where $\overline{\pi}_A=\sum_{i\in s_A} \pi_{iA}/n_A$.

In the case of a stratified design in both frames, let frame $A$ be divided into $H$ strata and let stratum $h$ has $N_{Ah}$ observation units of which $n_{Ah}$ are sampled. Similarly, frame $B$
has $L$ strata, the stratum $l$ has $N_{Bl}$ observation units  of which $n_{Bl}$ are sampled. Then, a Jackknife variance estimator of $\hat{Y}_{c}$ is given by
 \begin{equation} \label{eq:jackes}
v_J^{st} (\hat{Y}_{c}) =\sum_{h=1}^{H} \frac{n_{Ah}-1}{n_{Ah}}  \sum_{i\in s_{Ah}} (\hat{Y}_{c}^{A}(hi) -\overline{Y}_{c}^{Ah})^2
+ \sum_{l=1}^{L} \frac{n_{Bl}-1}{n_{Bl}} \sum_{j\in s_{Bl}} (\hat{Y}_{c}^{B}(lj)-\overline{Y}_{c}^{Bl})^2,
\end{equation}
where $\hat{Y}_{c}^{A}(hi)$ is the value taken by estimator  $\hat{Y}_{c}$ after dropping  unit $i$ of  stratum $h$ from sample $s_{Ah}$, $\overline{Y}_{c}^{Ah}$ is the average of these $n_{Ah}$ values; $\hat{Y}_{c}^{B}(lj)$ and $\overline{Y}_{c}^{Bl}$ are defined similarly. Again, we also can obtain a modified Jackknife variance estimator  $v_J^{st*} (\hat{Y}_{c})$ in stratified sampling using an approximate finite-population correction in each stratum. Asymptotic results for the Jackknife  estimators can be obtained using the approach presented in \citet{lohr2000inference}.

\section {Simulation studies}
\label{sec:sim}

We conduct an extensive simulation study to analyze the performance of the proposed estimators for surveys from two-frame finite populations. Our simulations are programmed in \texttt{R} using the \texttt{sampling} package developed by \cite{tille2006r} to draw the samples and to build all the calibration estimators, using the algorithms developed by \cite{wu2005algorithms} for the PEL approach and also developing some new \texttt{R}-code to compute calibration estimators with the Kullback-Leibler distance.

The simulated population has dimension $N=2350$. The values of the variable of interest $y$ are generated from a normal distribution $y_k\sim N(5000,500)$, for $k=1,\ldots,2350$. Units are randomly assigned to the two frames, $A$ and $B$, according to  three different scenarios depending on the overlap domain size $N_{ab}$. The first scenario has a \emph{small} overlap domain size and units are assigned to domain $a$, $b$ or $ab$ depending on the values taken by a binomial random variable $g_k \sim Bi(2,0.3)$. In particular, if $g_k=0$ then $k\in a$, if $g_k=1$ then $k\in b$ and if $g_k=2$ then $k\in ab$. The resulting sizes of the two frames are $N_A$=1309 and $N_B$=1251 and, consequently, the overlap domain size is $N_{ab}$=210.
The second and the third scenarios have respectively \emph{large} and \emph{medium} overlap domain size, depending on the values taken by $g_k \sim Bi(2,0.5)$, but assigning units to each domain in different ways for the two scenarios. In particular, we have $0$ for domain $a$, $1$ for domain $ab$ and $2$ for domain $b$ in the second scenario and $0$ for domain $b$, $1$ for domain $a$ and $2$ for domain $ab$ in the third scenario. The resulting frame sizes in the second scenario are given by $N_A$=1746 and $N_B$=1790 and the overlap domain size is $N_{ab}$=1186, while for the third scenario we have $N_A$=1790, $N_B$=1164 and $N_{ab}$=604.

Units from frame $A$ are then divided for each scenario into six strata as follows:
\begin{itemize}
\item sc.1 - large overlap, $N_{Ah}=(535,279,78, 148,101,168)$,
\item sc.2 - small overlap, $N_{Ah}=(734, 377, 116, 187, 115, 217)$,
\item sc.3 - medium overlap, $N_{Ah}=(781, 375, 114, 186, 111, 223)$.
\end{itemize}
Two auxiliary variables are then generated from the values of $y$ for frame $A$ and frame $B$, respectively, that are $x_{Ak}=(y_k-e_k)/0.5$ where $e_k \sim N(500,300)$ and $x_{B_k}=(y_k-1-e_k)/1.2$, where $e_k \sim N(700,500)$, for $k=1,\ldots,N$. The correlation coefficient with the variable of interest is given by $\rho_A=0.859$ and $\rho_B=0.709$, respectively.

Samples from frame $A$ are selected using stratified simple random sampling. Samples from frame $B$ are selected by means of Midzuno sampling, with inclusion probabilities proportional to variable $z_k = y_k-N(300,200)$, for $k=1,\ldots,N$ and having correlation $\rho=0.929$ with the  variable of interest. For each scenario, we draw four different combinations of sample sizes for frame $A$ and frame $B$, which correspond to the following number of units per stratum:
\begin{itemize}
\item $n_{A_{\ml{small}}}=(15,20,15,20,15,20)=105$ and $n_{B_{\ml{small}}}=135$,
\item $n_{A_{\ml{large}}}=(30,40,30,40,30,40)=210$ and $n_{B_{\ml{small}}}=135$,
\item $n_{A_{\ml{small}}}=(15,20,15,20,15,20)=105$ and $n_{B_{\ml{large}}}=270$,
\item $n_{A_{\ml{large}}}=(30,40,30,40,30,40)=210$ and $n_{B_{\ml{large}}}=270$.
\end{itemize}
This makes a $3\times 2 \times 2$ design for the simulation study. For each of the 12 settings, we compute four point calibration estimators of the population total $Y$ using both the single-frame and the dual-frame approach and using four different kinds of distance functions: Euclidean, Raking, Logit and Kullback-Leibler \citep[corresponding to the three methods considered in][and implemented in the \texttt{sampling package}, and the distance measure close to the PEL approach, respectively]{deville1993generalized}. For each  estimator we examine four different types of auxiliary information:
\begin{itemize}
\item[(1)] $N_A$, $N_B$, $N_{ab}$ all known,
\item[(2)] $N_A$, $N_B$ known and $N_{ab}$ unknown,
\item[(3)]{\color{black} $N_A$, $N_B$, $N_{ab}$, $X_{A}$, $X_{B}$  all known,}
\item[(4)]{\color{black} $N_A$, $N_B$, $X_{A}$, $X_{B}$  all known
and $N_{ab}$ unknown.}
\end{itemize}
We compute also the Hartley estimator (HAR), the Pseudo Maximum Likelihood estimator \citep[PML,][]{skinner1996estimation} when $N_{ab}$ is unknown, the single frame estimator \citep[SF,][]{bankier1986estimators,kalton1986sampling} and the Raking Ratio estimator \citep[SFRR,][]{skinner1991efficiency} for the purpose of comparison. When needed the value of $\eta$ has been estimated using
\begin{equation}\label{eq:etaest}
\hat{\eta}=N_aN_Bv(\hat{N}_{ba})/\left[N_bN_Av(\hat{N}_{ab})+N_aN_Bv(\hat{N}_{ba})\right],
\end{equation}
\citep[see][]{lohr2000inference}. {\color{black} Note that this choice allows for the computation of a single set of weights for all variables of interest. }For each estimator, we compute the percent relative bias $\m{RB\%}=E_{MC}(\hat{Y}-Y)/Y*100$, the percent relative mean squared error $\m{RMSE\%}={E_{MC}[(\hat{Y}-Y)^2]}/Y^2*100$ and the percent  gain in efficiency $\m{GE\%}=(1-\m{RMSE}/\m{RMSE}_{SF})*100$ over the single frame estimator $SF$, based on 1000 simulation runs.

Tables \ref{tab:sc1} to \ref{tab:sc3smallsmall} report results, one for each scenario, relative to the case in which $n_{A_{\ml{small}}}$ and $n_{B_{\ml{small}}}$, i.e. they are relatively smaller. The other cases are not reported since changing the sample size does not change the trend of the results. From  these tables we can see that relative biases are negligible in all cases, as expected from theoretical results. In terms of RMSE\%, other things being equal, single-frame estimators are more efficient than dual-frame estimators, and this can be explained by the extra-information they incorporate in the estimation process. Given a particular type of auxiliary information, it makes a little difference in terms of efficiency which distance metric we use in the calibration approach, and this is again in line with literature on the topic.

The performance in terms of efficiency of the estimators is essentially driven by the set of auxiliary variables employed, where type (3) -- $N_A$, $N_B$, $N_{ab}$, $X_{A}$, $X_{B}$  all known --  is the most effective as expected. In fact, the strong correlation between the study variable $y$ and the auxiliary variables $x_A$ and $x_B$ contributes in making estimates more accurate. It is interesting to note, however, that the performance of calibration estimators in setting (4) -- $N_A$, $N_B$, $X_{A}$, $X_{B}$  all known and $N_{ab}$ unknown -- is closer to that of setting (2) -- $N_A$, $N_B$ known and $N_{ab}$ unknown -- than that of setting (3), by this providing evidence of the importance of knowing the dimension of the overlap domain $N_{ab}$. This behavior becomes more clear as the overlap domain size becomes larger (Scenarios 3 and 2).

\begin{table}[ht] \small
\begin{center} 
\caption{Scenario 1: \emph{small} overlap domain size -- $N_{ab}$=210, $N_A$=1309, $N_B$=1251 -- sample sizes: $n_A$= 105, $n_B$= 135.} \label{tab:sc1}

\begin{tabular}{lcccccccc}  \hline \noalign{\smallskip}
                & &\multicolumn{3}{c}{\em  Single Frame}   &&\multicolumn{3}{c}{\em Dual Frame}                      \\ \noalign{\smallskip} \cline{3-5} \cline{7-9}  \noalign{\smallskip}
                & &{\sc rb \%}&{\sc 100*rmse \%}&{\sc ge \%}  &&{\sc rb \%}&{\sc 100*rmse \%}&{\sc ge \%} \\ \noalign{\smallskip} \cline{3-5} \cline{7-9}  \noalign{\smallskip} \noalign{\smallskip}

                                                                      & &\multicolumn{7}{c}{(1) \em $N_{A}, N_{ab}, N_{B}$ known}\\ \noalign{\smallskip} \noalign{\smallskip}
CAL (*)    & &-0.025 &  0.511 & 87.416  & &  -0.021 &  0.514 &  87.336 \\  \noalign{\smallskip}
                                                                      & &\multicolumn{7}{c}{(2) \em $N_{A}, N_{B}$ known, $N_{ab}$ unknown}\\ \noalign{\smallskip} \noalign{\smallskip}
HAR        & &   -   &   -    &    -    & &  -0.365 &  3.658 &   9.939 \\
PML        & &   -   &   -    &    -    & &   0.113 &  2.621 &  35.470 \\
SF         & &-0.147 &  4.062 &   0.000 & &    -    &    -   &     -   \\
SFRR       & &-0.128 &  2.315 &  43.002 & &    -    &    -   &     -   \\
CAL-EUC    & &-0.133 &  2.322 &  42.821 & &   0.032 &  2.587 &  36.313 \\
CAL-RAK    & &-0.128 &  2.315 &  43.002 & &   0.036 &  2.584 &  36.379 \\
CAL-LOG    & &-0.128 &  2.316 &  42.982 & &   0.035 &  2.584 &  36.372 \\
CAL-KL     & &-0.122 &  2.308 &  43.180 & &   0.039 &  2.581 &  36.443 \\ \noalign{\smallskip}
                                                                       & &\multicolumn{7}{c}{{\color{black}(3)} \em  $N_{A}, N_{ab}, N_{B}, {X}_{A}, {X}_{B}$ known}\\ \noalign{\smallskip}
CAL-EUC    & &-0.015 &  0.196 &  95.178 & &  -0.013 &  0.224 &  94.497 \\
CAL-RAK    & &-0.015 &  0.195 &  95.195 & &  -0.013 &  0.222 &  94.528 \\
CAL-LOG    & &-0.015 &  0.195 &  95.193 & &  -0.013 &  0.222 &  94.526 \\
CAL-KL     & &-0.013 &  0.195 &  95.199 & &  -0.010 &  0.224 &  94.481 \\
 \noalign{\smallskip} \noalign{\smallskip}\noalign{\smallskip} \noalign{\smallskip}
                                                                       & &\multicolumn{7}{c}{{\color{black}(4)} \em  $N_{A}, N_{B}, {X}_{A}, {X}_{B}$ known, $N_{ab}$ unknown}\\  \noalign{\smallskip}
CAL-EUC    & & 0.087 &  2.187 &  46.154 & &   0.087 &  2.231 &  45.083 \\
CAL-RAK    & & 0.086 &  2.186 &  46.181 & &   0.087 &  2.227 &  45.171 \\
CAL-LOG    & & 0.086 &  2.186 &  46.177 & &   0.087 &  2.227 &  45.162 \\
CAL-KL     & & 0.087 &  2.185 &  46.192 & &   0.088 &  2.229 &  45.132 \\  \noalign{\smallskip}
\hline
\multicolumn{9}{l}{\scriptsize (*) irrespective of the choice of the distance measure }
\end{tabular}
\end{center}
\end{table}

\begin{table}[ht] \small
\begin{center} 

\caption{Scenario 2: \emph{large} overlap domain size - $N_{ab}$=1186, $N_A$=1746, $N_B$=1790. Samples sizes: $n_A$= 105, $n_B$= 135} \label{tab:sc2smallsmall}
\begin{tabular}{lcccccccc} \hline \hline \noalign{\smallskip}
                & &\multicolumn{3}{c}{\em  Single Frame}   &&\multicolumn{3}{c}{\em Dual Frame}                      \\ \noalign{\smallskip} \cline{3-5} \cline{7-9}  \noalign{\smallskip}
                & &{\sc rb \%}&{\sc 100*rmse \%}&{\sc ge \%}  &&{\sc rb \%}&{\sc 100*rmse \%}&{\sc ge \%} \\ \noalign{\smallskip} \cline{3-5} \cline{7-9}  \noalign{\smallskip} \noalign{\smallskip}
                                                               & &\multicolumn{7}{c}{(1) \em $N_{A}, N_{ab}, N_{B}$ known}\\ \noalign{\smallskip} \noalign{\smallskip}
CAL (*)    & & 0.021 & 0.578   & 95.282  & &    0.026 &  0.605 &  95.059 \\

                                                                         & &\multicolumn{7}{c}{(2) \em $N_{A}, N_{B}$ known, $N_{ab}$ unknown}\\ \noalign{\smallskip} \noalign{\smallskip}
HAR        & &   -   &   -     &    -    & &   -0.144 &  9.399 &  23.252 \\
PML        & &   -   &   -     &    -    & &   -0.029 &  7.505 &  38.717 \\
SF         & &-0.062 &  12.246 & 0.000   & &     -    &    -   &    -    \\
SFRR       & &-0.219 &   7.251 & 40.788  & &     -    &    -   &    -    \\
CAL-EUC    & &-0.327 &   7.448 & 39.178  & &   -0.336 &  7.765 &  36.591 \\
CAL-RAK    & &-0.219 &   7.251 & 40.788  & &   -0.240 &  7.572 &  38.170 \\
CAL-LOG    & &-0.231 &   7.271 & 40.624  & &   -0.250 &  7.591 &  38.009 \\
CAL-KL     & &-0.108 &   7.114 & 41.905  & &   -0.143 &  7.426 &  39.359 \\ \noalign{\smallskip} \noalign{\smallskip}
                                                                         & &\multicolumn{7}{c}{{\color{black}(3)} \em  $N_{A}, N_{ab}, N_{B}, {X}_{A}, {X}_{B}$ known}\\ \noalign{\smallskip} \noalign{\smallskip}
CAL-EUC    & & 0.026 &   0.215 & 98.241  & &    0.034 &  0.301 &  97.542 \\
CAL-RAK    & & 0.027 &   0.215 & 98.244  & &    0.035 &  0.298 &  97.563 \\
CAL-LOG    & & 0.027 &   0.215 & 98.244  & &    0.035 &  0.299 &  97.562 \\
CAL-KL     & & 0.030 &   0.239 & 98.047  & &    0.038 &  0.312 &  97.451 \\ \noalign{\smallskip} \noalign{\smallskip}
                                                                         & &\multicolumn{7}{c}{{\color{black}(4)} \em  $N_{A}, N_{B}, {X}_{A}, {X}_{B}$ known, $N_{ab}$ unknown}\\ \noalign{\smallskip} \noalign{\smallskip}
CAL-EUC    & &-0.278 &   7.259 & 40.721  & &   -0.271 &  7.353 &  39.956\\
CAL-RAK    & &-0.278 &   7.261 & 40.706  & &   -0.270 &  7.353 &  39.958\\
CAL-LOG    & &-0.278 &   7.261 & 40.707  & &   -0.270 &  7.353 &  39.956\\
CAL-KL     & &-0.278 &   7.263 & 40.687  & &   -0.274 &  7.360 &  39.895\\  \noalign{\smallskip} \noalign{\smallskip}
\hline
\multicolumn{9}{l}{\scriptsize (*) irrespective of the choice of the distance measure }
\end{tabular}
\end{center}
\end{table}

\begin{table}[ht] \small
\begin{center} 
\caption{Scenario 3: \emph{medium} overlap domain size -
$N_{ab}$=604, $N_A$=1790, $N_B$=1164 Samples sizes: $n_A$= 105,
$n_B$= 135} \label{tab:sc3smallsmall}
\begin{tabular}{lcccccccc} \hline \hline \noalign{\smallskip}
                & &\multicolumn{3}{c}{\em  Single Frame}   &&\multicolumn{3}{c}{\em Dual Frame}                      \\ \noalign{\smallskip} \cline{3-5} \cline{7-9}  \noalign{\smallskip}
                & &{\sc rb \%}&{\sc 100*rmse \%}&{\sc ge \%}  &&{\sc rb \%}&{\sc 100*rmse \%}&{\sc ge \%} \\ \noalign{\smallskip} \cline{3-5} \cline{7-9}  \noalign{\smallskip} \noalign{\smallskip}
                                                               & &\multicolumn{7}{c}{(1) \em $N_{A}, N_{ab}, N_{B}$ known}\\ \noalign{\smallskip} \noalign{\smallskip}
CAL (*)    & & 0.006&  0.761 &  95.036 & &   0.007 &  0.779 &  94.920 \\ \noalign{\smallskip} \noalign{\smallskip}
                                                                      & &\multicolumn{7}{c}{(2) \em $N_{A}, N_{B}$ known, $N_{ab}$ unknown}\\ \noalign{\smallskip} \noalign{\smallskip}
HAR        & &   -  &   -    &    -    & &  -0.016 & 13.453 &  12.268 \\
PML        & &   -  &   -    &    -    & &   0.265 &  4.513 &  70.567 \\
SF         & &0.213 & 15.334 &  0.000  & &    -    &    -   &    -    \\
SFRR       & &0.055 &  4.271 &  72.148 & &    -    &    -   &    -    \\
CAL-EUC    & &0.013 &  4.333 &  71.744 & &   0.074 &  4.510 &  70.587 \\
CAL-RAK    & &0.055 &  4.271 &  72.148 & &   0.109 &  4.469 &  70.855 \\
CAL-LOG    & &0.050 &  4.277 &  72.108 & &   0.105 &  4.473 &  70.828 \\
CAL-KL     & &0.096 &  4.230 &  72.417 & &   0.144 &  4.442 &  71.032 \\ \noalign{\smallskip} \noalign{\smallskip}
                                                                      & &\multicolumn{7}{c}{{\color{black}(3)} \em  $N_{A}, N_{ab}, N_{B}, {X}_{A}, {X}_{B}$ known}\\ \noalign{\smallskip} \noalign{\smallskip}
CAL-EUC    & &0.004 &  0.226 &  98.527 & &  -0.006 &  0.356 &  97.676 \\
CAL-RAK    & &0.002 &  0.226 &  98.526 & &  -0.007 &  0.354 &  97.694 \\
CAL-LOG    & &0.002 &  0.226 &  98.526 & &  -0.007 &  0.354 &  97.693 \\
CAL-KL     & &0.000 &  0.227 &  98.523 & &  -0.008 &  0.355 &  97.683 \\ \noalign{\smallskip} \noalign{\smallskip}
                                                                      & &\multicolumn{7}{c}{{\color{black}(4)} \em  $N_{A}, N_{B}, {X}_{A}, {X}_{B}$ known, $N_{ab}$ unknown}\\ \noalign{\smallskip} \noalign{\smallskip}
CAL-EUC    & &0.174 &  3.762 &  75.468 & &   0.163 &  3.955 &  74.207  \\
CAL-RAK    & &0.172 &  3.762 &  75.465 & &   0.163 &  3.956 &  74.200  \\
CAL-LOG    & &0.172 &  3.762 &  75.465 & &   0.163 &  3.956 &  74.201  \\
CAL-KL     & &0.170 &  3.762 &  75.468 & &   0.163 &  3.974 &  74.082  \\ \noalign{\smallskip} \noalign{\smallskip}
\multicolumn{9}{l}{\scriptsize (*) irrespective of the choice of the distance measure }
\end{tabular}
\end{center}
\end{table}

As discussed above, if we consider the calibration estimator with the Kullback-Leiber distance (CAL-KL) and we add the constraint induced by the common domain mean as in equation (\ref{PEL-constr-dualframe2}), we obtain the PEL estimator proposed by \citet{rao2010pseudo}. To evaluate the effect of including such restriction  in the calibration process, we have also computed the CAL-EUC and the CAL-KL  estimators that include this new restriction (overlap restriction) in all scenarios and under two particular types of auxiliary information: (1) and (3). Table \ref{tab:extra-constr} reports the results from this experiment. It can be noted that in this simulation study,  the inclusion of this extra constraint provides little or no improvement over classical calibration. In particular, in case (1) calibration estimators without restriction work a little better than the estimators that include the restriction, while in case (3) this behavior is reversed. {\color{black}Note also that using this extra constraint comes at the price of having a final estimator non-linear in $y$ and, therefore, would require different sets of weights for different variables of interest. Therefore, when used in large scale surveys one may want to choose a subset of variables of interest to enter such extra benchmark constraints and then use the final set of weights for all computations.}

\begin{table}[ht] \small
\begin{center} 
\caption{Efficiency of Kullback-Leiber ({\sc kl}) and Euclidean ({\sc euc}) distance based calibration estimators \emph{With} and \emph{Without} overlap restriction (\ref{PEL-constr-dualframe2}) in the dual frame approach.}\label{tab:extra-constr}
\begin{tabular}{llcccccccc} \hline \hline \noalign{\smallskip}
                                                                                        &&&\multicolumn{3}{c}{\em  With}   &&\multicolumn{3}{c}{\em Without}                      \\ \noalign{\smallskip} \cline{4-6} \cline{8-10}  \noalign{\smallskip}
                                                                                        \emph{Scenario}&$(n_A, n_B)$&&{\sc rb \%}&{\sc 100*rmse \%}&{\sc ge \%}  &&{\sc rb \%}&{\sc 100*rmse \%}&{\sc ge \%} \\ \noalign{\smallskip} \cline{4-6} \cline{8-10}  \noalign{\smallskip} \noalign{\smallskip}
                                                                                        &&&\multicolumn{7}{c}{(1) \em $N_{A}, N_{B}$, $N_{ab}$,  known}\\ \noalign{\smallskip} \noalign{\smallskip}
\emph{Small} & (105,135) &{\sc kl} &   -0.019 & 0.529 & 86.972 & &  -0.021 & 0.514 & 87.336 \\ \noalign{\smallskip}
             &           &{\sc euc} &   -0.020 & 0.521 & 87.171 & &  -0.021 & 0.514 & 87.336 \\ \noalign{\smallskip}\noalign{\smallskip}
             & (210,270) &{\sc kl} &    0.017 & 0.253 & 87.260 & &   0.020 & 0.244 & 87.709 \\ \noalign{\smallskip}
             &           &{\sc euc} &    0.020 & 0.249 & 87.446 & &   0.020 & 0.244 & 87.709 \\ \noalign{\smallskip}\noalign{\smallskip}
\emph{Large} & (105,135) &{\sc kl} &    0.021 & 0.644 & 94.742 & &   0.026 & 0.605 & 95.059 \\ \noalign{\smallskip}
             &           &{\sc euc} &    0.025 & 0.643 & 94.750 & &   0.026 & 0.605 & 95.059 \\ \noalign{\smallskip}\noalign{\smallskip}
             & (210,270) &{\sc kl} &   -0.004 & 0.273 & 95.224 & &  -0.006 & 0.258 & 95.487 \\ \noalign{\smallskip}
             &           &{\sc euc} &   -0.003 & 0.274 & 95.211 & &  -0.006 & 0.258 & 95.487 \\ \noalign{\smallskip}\noalign{\smallskip}
\emph{Medium}& (105,135) &{\sc kl} &    0.022 & 0.817 & 94.672 & &   0.007 & 0.779 & 94.920 \\ \noalign{\smallskip}
             &           &{\sc euc} &    0.017 & 0.813 & 94.698 & &   0.007 & 0.779 & 94.920 \\ \noalign{\smallskip}\noalign{\smallskip}
             & (210,270) &{\sc kl} &   -0.005 & 0.385 & 94.341 & &  -0.002 & 0.367 & 94.606 \\ \noalign{\smallskip}
             &           &{\sc euc} &   -0.004 & 0.385 & 94.344 & &  -0.002 & 0.367 & 94.606 \\ \noalign{\smallskip}\noalign{\smallskip} \noalign{\smallskip}
                                                                                                   &&&\multicolumn{7}{c}{(3) \em  $N_{A}, N_{ab}, N_{B}, {X}_{A}, {X}_{B}$ known}\\ \noalign{\smallskip} \noalign{\smallskip}  \noalign{\smallskip}
\emph{Small} & (105,135) &{\sc kl} &   -0.015 & 0.215 & 94.711 & &  -0.013 & 0.224 & 94.481 \\ \noalign{\smallskip}
             &           &{\sc euc} &   -0.016 & 0.209 & 94.844 & &  -0.010 & 0.224 & 94.497 \\ \noalign{\smallskip}\noalign{\smallskip}
             & (210,270) &{\sc kl} &    0.010 & 0.094 & 95.242 & &   0.015 & 0.100 & 94.960 \\ \noalign{\smallskip}
             &           &{\sc euc} &    0.011 & 0.094 & 95.279 & &   0.015 & 0.101 & 94.914 \\ \noalign{\smallskip}\noalign{\smallskip}
\emph{Large} & (105,135) &{\sc kl} &    0.025 & 0.306 & 97.501 & &   0.034 & 0.312 & 97.451 \\ \noalign{\smallskip}
             &           &{\sc euc} &    0.031 & 0.278 & 97.731 & &   0.038 & 0.301 & 97.542 \\ \noalign{\smallskip}\noalign{\smallskip}
             & (210,270) &{\sc kl} &   -0.004 & 0.121 & 97.883 & &  -0.001 & 0.134 & 97.663 \\ \noalign{\smallskip}
             &           &{\sc euc} &   -0.003 & 0.121 & 97.878 & &  -0.001 & 0.135 & 97.643 \\ \noalign{\smallskip}\noalign{\smallskip}
\emph{Medium}& (105,135) &{\sc kl} &    0.008 & 0.292 & 98.097 & &  -0.006 & 0.355 & 97.683 \\ \noalign{\smallskip}
             &           &{\sc euc} &    0.010 & 0.276 & 98.197 & &  -0.008 & 0.356 & 97.676 \\ \noalign{\smallskip}\noalign{\smallskip}
             & (210,270) &{\sc kl} &    0.001 & 0.128 & 98.120 & &   0.003 & 0.173 & 97.465 \\ \noalign{\smallskip}
             &           &{\sc euc} &    0.003 & 0.128 & 98.119 & &   0.003 & 0.174 & 97.448 \\ \noalign{\smallskip}\noalign{\smallskip}
\end{tabular}
\end{center}
\end{table}

We now turn to the construction of confidence intervals for $Y$.  We obtain the  $95\%$ confidence intervals based on a normal distribution and the two proposed variance estimators: linearization based $v(\hat{Y}_{GREG})$ from Theorem \ref{the:estvar} and Corollary \ref{cor:cal} and Jackknife as in equation (\ref{eq:jackes}) with finite-population correction. Table \ref{tab:var}  shows the average length of $95\%$ confidence intervals, the empirical  coverage probability, the inferior and the superior tail error rates.  For space reason, only some cases and some sample sizes are included.

From Table \ref{tab:var} we can observe that   coverage probability is high (greater than $93\%$) for all sample sizes and all scenarios. We also observe that the intervals based on the linearization variance tend to provide empirical coverages larger than the nominal ones, while Jackknife based intervals provide coverage  closer to the nominal. Jackknife intervals are also  shorter and the length difference is significant in some cases (e.g. large overlap size  together with the $X_A$ and $X_B$ information). The worse performance  of the linearization based  intervals may be due to the fact that sample sizes in some strata are too small.

\begin{table}[ht] \small
\begin{center}
\caption{ Length $95\%$ confidence interval, inferior and superior tail error rate, empirical coverage. Linearization and Jackknife variance estimators of the CAL-EUC estimators.}\label{tab:var}
\begin{tabular}{lcccccccccc}  \hline \noalign{\smallskip}
Linearization/   &&\multicolumn{4}{c}{\em  Single Frame}   &&\multicolumn{4}{c}{\em Dual Frame}                       \\ \noalign{\smallskip} \cline{3-6} \cline{8-11}  \noalign{\smallskip}
Jackknife     & $(n_A, n_B)$ &{\sc len}&{\sc inf \%}&{\sc sup \%}&{\sc cov \%}  &&{\sc len}&{\sc inf \%}&{\sc sup \%}&{\sc cov \%} \\ \noalign{\smallskip} \cline{3-6} \cline{8-11}  \noalign{\smallskip} \noalign{\smallskip}
                                                                            \multicolumn{2}{l}{\em Sc.1: Small overlap size}&\multicolumn{9}{c}{\em  $N_{A}, N_{ab}, N_{B}$ known}\\ \noalign{\smallskip} \noalign{\smallskip}
Lin  &(105,135) & 360067 & 1.6  &  2.1  & 96.3 & & 365977 & 1.4  & 2.3  & 96.3 \\
Jack &          & 337968 &   2.1&   3.10&  94.8& &  341008&   2.2&   2.6&  95.2\\ \noalign{\smallskip}\noalign{\smallskip}
Lin  &(210,270) & 243142 & 2.5  &  1.5  & 96.0 & & 249887 & 1.9  & 1.3  & 96.8 \\
Jack &          & 233017 &   2.7&   2.00&  95.3& &  235151&   2.8&   1.8&  95.4\\ \noalign{\smallskip} \noalign{\smallskip}
                                                                               & &\multicolumn{9}{c}{\em  $N_{A}, N_{ab}, N_{B}, {X}_{A}, {X}_{B}$ known}\\ \noalign{\smallskip} \noalign{\smallskip}
Lin  &(105,135) & 295610 & 0.1  &  0.6  & 99.3 & & 312865 & 0.1  & 1.0  & 98.9 \\
Jack &          & 201311 &   2.4&   3.10&  94.5& &  203118&   2.8&   2.9&  94.3\\ \noalign{\smallskip}\noalign{\smallskip}
Lin  &(210,270) & 192396 & 1.3  &  0.6  & 98.1 & & 212015 & 0.2  & 0.1  & 99.7 \\
Jack &          & 137885 &   3.5&   1.90&  94.6& &  139089&   3.2&   1.8&  95.0\\ \noalign{\smallskip} \noalign{\smallskip}
                                                                               \multicolumn{2}{l}{\em Sc.2: Large overlap size}&\multicolumn{9}{c}{\em  $N_{A}, N_{ab}, N_{B}$ known}\\ \noalign{\smallskip} \noalign{\smallskip}
Lin  &(105,135) & 458024 & 1.9  & 0.5  & 97.6 & & 513298 & 0.3  & 0.4  & 99.3 \\
Jack &          & 344233 &   3.6&   2.9&  93.5& &  376839&   3.8&   2.8&  93.4\\ \noalign{\smallskip}\noalign{\smallskip}
Lin  &(210,270) & 292011 & 2.3  & 1.4  & 96.3 & & 356127 & 0.5  & 0.1  & 99.4 \\
Jack &          & 237054 &   2.3&   1.9&  95.8& &  258610&   2.2&   2.3&  95.5\\ \noalign{\smallskip} \noalign{\smallskip}
                                                                               & &\multicolumn{9}{c}{\em  $N_{A}, N_{ab}, N_{B}, {X}_{A}, {X}_{B}$ known}\\ \noalign{\smallskip} \noalign{\smallskip}
Lin  &(105,135) & 395270 & 0.7  & 1.0  & 98.3 & & 441181 & 0.1  & 0.1  & 99.8 \\
Jack &          & 207237 &   3.2&   2.8&  94.0& &  220257&   2.8&   2.7&  94.5\\ \noalign{\smallskip}\noalign{\smallskip}
Lin  &(210,270) & 268575 & 1.2  & 0.9  & 97.9 & & 303252 & 0.0  & 0.0  & 100. \\
Jack &          & 141342 &   2.6&   2.5&  94.9& &  149709&   2.2&   2.7&  95.1\\ \noalign{\smallskip} \noalign{\smallskip}
                                                                               \multicolumn{2}{l}{\em  Sc.3: Medium overlap size}&\multicolumn{9}{c}{\em $N_{A}, N_{ab}, N_{B}$ known}\\ \noalign{\smallskip} \noalign{\smallskip}
Lin  &(105,135) & 459903 & 1.9  & 1.9  & 96.2 & & 485557 & 1.7  & 1.3  & 97.0 \\
Jack &          & 394059 &   3.8&   2.7&  93.5& &  400930&   4.1&   2.2&  93.7\\ \noalign{\smallskip}\noalign{\smallskip}
Lin  &(210,270) & 314875 & 1.9  & 1.9  & 96.2 & & 336503 & 0.9  & 1.3  & 97.8 \\
Jack &          & 276678 &   2.5&   3.1&  94.4& &  280932&   2.6&   3.4&  94.0\\ \noalign{\smallskip} \noalign{\smallskip}
                                                                              & &\multicolumn{9}{c}{\em  $N_{A}, N_{ab}, N_{B}, {X}_{A}, {X}_{B}$ known}\\ \noalign{\smallskip} \noalign{\smallskip}
Lin &(105,135) & 312413 & 1.6  & 1.6  & 96.8 & & 370388 & 0.9  & 0.2  & 98.9  \\
Jack &          & 216987 &   2.4&   3.0&  94.6& &  220173&   2.3&   2.1&  95.6\\ \noalign{\smallskip}\noalign{\smallskip}
Lin &(210,270) & 189223 & 3.2  & 2.8  & 94.0 & & 251778 & 0.7  & 0.5  & 98.8  \\
Jack &          & 149740 &   2.7&   2.1&  95.2& &  151651&   2.5&   1.9&  95.6\\ \noalign{\smallskip} \noalign{\smallskip}\hline
\end{tabular}
\end{center}
\end{table}

\section{Application}\label{sec:appl}

IESA, the Institute for Advanced Social Studies of Spain conducted a survey between  January, 14th  and February, 13th 2011 on the perception of culture in the Spanish region of Andalusia
(Barometer of Culture of Andalusia - BACU). It is based on a sample drawn from two frames: landline phone frame ($A$, $N_{A}=$ 5,064,304) and a mobile phone frame
($B$, $N_{B}=$ 5,875,280). The overlap domain size is known to have dimension $N_{ab}=$ 4,421,042.

From frame $A$ a stratified random sample without replacement of dimension $n_{A}=641$ was selected, where strata are made by eight geographical regions.
Strata population sizes  in frame $A$ are $N_{Ah}$ = (274128, 919124, 463008, 502450, 237183, 441936, 856392,  1370083) and the corresponding strata sample sizes
are $n_{Ah}$ = (53,  99,  66,  62,  38,  49, 131, 143). From frame $B$ a simple random sample without replacement of size $n_{B}=177$ was drawn.
Sample sizes for each frame were determined  so as to minimize the cost of the survey.

Among the several topics of interest in the survey,  there is also the interest to estimate the percentage of undecided  citizens on next political elections.  As auxiliary variables there are available  sex and age (in two categories, under 45 or over); both variables are observed in both frames and their totals are known for each of the two frames $A$ and $B$.

We compare estimates of the mean of such binary variable of interest without using any auxiliary information, using the auxiliary information provided by the sizes of the frames and overlap domains, and also using additional auxiliary information from age and sex. Results are reported in Table \ref{tab:ex}. In particular, without auxiliary information, under the dual frame approach,  we compute
the Hartley estimator (HAR) estimating  $\eta$ as in (\ref{eq:etaest})
and, under the single frame approach, we compute Kalton-Anderson's (SF) estimator. Dual frame pseudo-maximum likelihood (PML) estimator and the single frame raking ratio estimation
(SFRR) are also computed. Calibration estimators using two
levels of auxiliary information and four distances, are also reported in Table \ref{tab:ex}. The confidence intervals (and their length) based on Jackknife
variance estimation are included as well.

From Table \ref{tab:ex} we observe that the inclusion of auxiliary information provides estimates with shorter confidence intervals. This is particularly true for when using calibration on population domains, sex and age under the single frame approach. Calibration estimates (including SFRR) are all similar, and this is particularly true when comparing values within the single and the dual frame framework. However, including all available auxiliary information, both in terms of the design -- hence using the single frame approach -- and of population counts we obtain the best empirical performance and an estimate that is, nonetheless, coherent with the others.

\begin{table}[ht]
\begin{center}
\caption{Estimated proportion ($\hat{\hbox{\sc p}}$), lower bound ({\sc lb}), upper bound ({\sc ub}) and length ({\sc l}) of a 95\% confidence interval
under dual and single frame approach for alternative  estimators}\label{tab:ex}
\begin{tabular}{lcccccccccc} \hline \hline \noalign{\smallskip}
          &&\multicolumn{4}{c}{\em Single Frame} &&\multicolumn{4}{c}{\em Dual Frame} \\ \noalign{\smallskip} \cline{3-6} \cline{8-11} \noalign{\smallskip}
          &&$\hat{\hbox{\sc p}}$\%&{\sc lb}&{\sc ub}&{\sc l} &&$\hat{\hbox{\sc p}}\%$&{\sc lb}&{\sc ub}&{\sc l} \\ \noalign{\smallskip} \cline{3-6} \cline{8-11} \noalign{\smallskip} \noalign{\smallskip}
{HAR    } &&      - &    - &     - &    - &&  9.03 & 6.10 & 11.95 & 5.85\\
{SF     } &&  11.61 & 8.68 & 14.55 & 5.88 &&     - &    - &     - &    -\\
{PML }    &&      - &    - &     - &    - && 11.28 & 7.15 & 15.40 & 8.26\\
{SFRR}    &&  11.25 & 8.81 & 13.70 & 4.89 &&     - &    - &     - &    -\\ \noalign{\smallskip} & &\multicolumn{9}{c}{\em $N_{a}, N_{ab}, N_{b}$ known}
                                                                        \\ \noalign{\smallskip} \noalign{\smallskip}
{CAL (*)} &&  10.97 & 8.68 & 13.27 & 4.60 &&  9.49 & 7.08 & 11.90 & 4.82\\ \noalign{\smallskip} \noalign{\smallskip} & &\multicolumn{9}{c}{\em  $N_{a}, N_{ab}, N_{b}, X_{A}, X_{B}$ known}
                                                                        \\ \noalign{\smallskip} \noalign{\smallskip}
{CAL-EUC} &&  10.73 & 8.51 & 12.95 & 4.43 &&  9.06 & 6.72 & 11.40 & 4.67\\
{CAL-RAK} &&  10.76 & 8.52 & 12.99 & 4.47 &&  9.12 & 6.76 & 11.48 & 4.72\\
{CAL-LOG} &&  10.76 & 8.53 & 12.99 & 4.46 &&  9.11 & 6.75 & 11.47 & 4.71\\
{CAL-KL } &&  10.71 & 8.47 & 12.95 & 4.48 &&  9.22 & 6.78 & 11.66 & 4.88\\ \noalign{\smallskip} \noalign{\smallskip} \hline
\multicolumn{11}{l}{\scriptsize (*) irrespective of the choice of the distance measure }
\end{tabular}
\end{center}
\end{table}

\section{Conclusions}\label{sec:concl}

In the last years multiple-frame surveys have significantly attracted attention in survey methodology and applications. The use of more then one frame helps statisticians to obtain more reliable estimates for finite population totals or means. Incorporating available auxiliary
population information at different levels also contributes to obtain more
accurate estimates. In this work we have discussed the extension of the calibration framework to estimation from dual frame surveys. Definition of the auxiliary variables and benchmark constraints have been discussed under both the single and the dual frame approach. Some of the estimators already proposed in the literature have been shown to belong to this class of calibration estimators. {\color{black} The cases discussed in Section 3 are only a few examples of the very many possible ones that can be treated with calibration. The calibration approach is very flexible and wide spread for one frame surveys. We wanted to import such flexibility in the field of two frame surveys.}

Estimators belonging to this class have been proven to be design consistent under mild assumptions and their asymptotic distribution has been obtained. Variance estimation has been proposed under the linearization and the Jackknife framework. Results from the extensive simulation study support theoretical findings and show that, given a set of auxiliary variables, the choice of a distance function  makes little difference in terms of efficiency, as it is the case also in one frame surveys. {\color{black}In addition, it is well known that   calibration based on the Euclidean distance function can produce negative weights whilst  calibration based on the Kullback-Leibler divergence or on  other distance functions considered in this paper ensures always positive weights. In this paper, we have found that in the application, the calibration estimator with Euclidean distance does not give negative weights. In the simulation study, on the other hand, among the 24,000 samples (4 cases $\times$ 3 scenarios $\times$ 2 sample sizes $\times$ 1000 replicates), the calibration estimator with Euclidean distance gives negative weights in only 6 cases in Single Frame (under scenario 1, with a relatively smaller sample size and with auxiliary information of type $(3)$ and $(4)$), while DF in 981 cases (in all scenarios, when with auxiliary information of type $(4)$).}

Calibration estimation from dual frame surveys can be implemented easily using existing software for one frame populations, as for the application on data from the BACU survey. Note that calibration weights can be applied to all variables of interest. {\color{black} In fact, they do not depend on the value taken by the  variable of interest. This is particularly valuable, because in this way calibration estimators  give internal consistency.  With repeated surveys, the simplicity and transparency of a fixed-weight estimator may be preferred. Fixed-weight adjustments may make year-to-year comparisons easier in an annual survey, where the domain proportions are relatively constant over time. Standard survey software may then be used to estimate population totals using the modified weights.}

{\color{black} The proposed calibration estimators assume that the control totals are values known without sampling errors. However, these control totals can themselves be estimated from other surveys. Calibration can be applied similarly with those estimated controls but in this case the variance estimator need to take into account such extra variation when we use estimates of totals. To obtain the variance estimator when the controls are estimated a possibility is to use the result of Section 9 in \citet{berger2009variance} for each sample $s_A$ and $s_B$ separately. These Authors obtain a variance estimator of the calibration estimator that takes into account the randomness of multiple estimates controls.}

The extension to more than two frames is under study as well. {\color{black}One important issue when dealing with more than two frames is that of using a proper notation \citep[see][]{lohr2006estimation,mecatti2009generalized}}. A first simple way around is the one, also considered in \citet{rao2010pseudo}, in which weights from the multiplicity estimator of \citet{mecatti2007single} are used as starting weights and calibration is applied straightforwardly.  More complicated is the issue of accounting for different levels of  frame information, although we believe that \citet{mecatti2009generalized} may provide a good starting point.  {\color{black}{In addition, note that, given that with calibration estimation we are often estimating totals in domains using ratio type estimators (like with post-stratification), the sample size of the domains is important to avoid the introduction of possible bias in the final estimates. This issue becomes particularly relevant when moving to more than two frames. In this case, domains may easily become small areas and model based techniques could be enforced to fully exploit auxiliary information. }}

\section*{Acknowledgements} The Authors are grateful to Manuel Trujillo (IESA) for providing  data and information about the Barometer of Culture of Andalusia survey and to Jean-Claude Deville for useful suggestions on distance metics in calibration. This Research is partially supported by Ministerio de Educaci\'on y Ciencia (grant MTM2012-35650, Spain) and by Consejer\'ia de Econom\'ia, Innovaci\'on, Ciencia y Empleo (grant SEJ2954, Junta de Andaluc\'ia). The work of Ranalli has been developed partially under the support of the project PRIN-SURWEY (grant 2012F42NS8, Italy).

\appendix
\numberwithin{equation}{section}
\section{Proofs}

\textbf{Proof of Theorem \ref{the:cons}} \\
By assumptions A\ref{as:B} and A\ref{as:poscov} we have
\begin{align*}
\hat{Y}_{\ml{GREG}}-Y & =\hat{Y}_H+(\b{t}_x-\ha{t}_{xH})\b{B}_U-Y+(\b{t}_x-\ha{t}_{xH})(\ha{\beta}^{\circ}-\b{B}_U)\nonumber\\
&=\hat{Y}_H+(\b{t}_x-\ha{t}_{xH})\b{B}_U-Y+O_p(Nn_N^{-1/2})o_p(1).
\end{align*}
Now, $\hat{Y}_H+(\b{t}_x-\ha{t}_{xH})\b{B}_U$ is such that a central limit theorem holds for A\ref{as:poscov} and A\ref{as:clt}, i.e.
\begin{align*}
\frac{\sqrt{n_N}}{N}(\hat{Y}_H+(\b{t}_x-\ha{t}_{xH})\b{B}_U-Y)\to^{\hskip -0.125in \mathcal{L}}N(0,\nu^2)
\end{align*}
where $\nu^2=\Sigma_{yy}-2\b{\Sigma}_{xy}\b{B}+\b{B}^{T}\b{\Sigma}_{xx}\b{B}$. Now, $N^2n_NV(\hat{t}_{eH})\rightarrow \nu^2$ as $N\rightarrow\infty$, so that $\hat{Y}_H+ (\b{t}_x-\ha{t}_{xH})\b{B}_U-Y=O_p(Nn_N^{-1/2})$ and the result follows.

\qed

~\\

\noindent\textbf{Proof of Theorem \ref{the:estvar}} \\
Let $\tilde{y}_k=\b{x}_k\b{B}_U$ and $\hat{y}_k=\b{x}_k\ha{\beta}^{\circ}$. Then
\begin{align}
v(\hat{t}_{\hat{e}H})&=v(\hat{t}_{\hat{e}H}+\hat{t}_{eH}-\hat{t}_{eH})=\nonumber\\
&=v(\sum_{k \in s}d_k^{\circ}\hat{e}_k+\sum_{k \in s}d_k^{\circ}e_k-\sum_{k \in s}d_k^{\circ}e_k)=\nonumber\\
&=v\Big(\sum_{k \in s}d_k^{\circ}e_k+\sum_{k \in s}d_k^{\circ}(y_k-\hat{y}_k-y_k+\tilde{y}_k)\Big)=\nonumber\\
&=v(\hat{t}_{eH})+v(\hat{t}_{\tilde{y}-\hat{y},H})+2c(\hat{t}_{eH},\hat{t}_{\tilde{y}-\hat{y},H}).
\end{align}

Now, for A\ref{as:B}, A\ref{as:poscov} and A\ref{as:estvar}, we have
\begin{enumerate}
  \item $v(\hat{t}_{eH})=V(\hat{t}_{eH})+o_p(N^2n_N^{-1})$,
  \item
      $v(\hat{t}_{\tilde{y}-\hat{y},H})=v(\sum_{k \in s}d_k^{\circ}\b{x}_k(\b{B}_U-\ha{\beta}^{\circ}))=(\b{B}_U-\ha{\beta}^{\circ})^{T}v(\ha{t}_{xH})
      (\b{B}_U-\ha{\beta}^{\circ})
      =o_p(1)O_p(N^2n_N^{-1})o_p(1)$,
  \item $c(\hat{t}_{eH},\hat{t}_{\tilde{y}-\hat{y},H})=c\big(\sum_{k \in s}d_k^{\circ}e_k,\sum_{k \in s}d_k^{\circ}\b{x}_k(\b{B}_U-\ha{\beta}^{\circ})\big)=\b{c}(\hat{t}_{eH},\ha{t}_{xH})(\b{B}_U-\ha{\beta}^{\circ})
       =O_p(N^2n_N^{-1})o_p(1)$.
\end{enumerate}
\qed

\noindent\textbf{Proof of Theorem \ref{the:cal}} \\
Using Result 3 in \citet{Deville1992calibration}
$$
\b{\lambda}=\Big(\sum_{k \in s}d_k^{\circ}\,\b{x}_k^T\b{x}_k\Big)^{-1}(\b{t}_x-\ha{t}_{xH})^{T}+O_p(n_N^{-1}),
$$
$w_k=d_k^{\circ}F(\b{x}_k\b{\lambda})=:d_k^{\circ}(1+\b{x}_k\b{\lambda})+\epsilon_k(\b{x}_k\b{\lambda})$. Assumption A\ref{as:boundx} ensures that $\epsilon_k(u)=O_p(u^2)$, therefore
$$
\hat{Y}_{\ml{CAL}}=\hat{Y}_{\ml{GREG}}+O_p(Nn_N^{-1})+O_p(Nn_N^{-2}).
$$
\qed

\bibliography{biblio}

\end{document}